\documentclass[10pt,conference]{IEEEtran}
\IEEEoverridecommandlockouts

\usepackage[T1]{fontenc}
\usepackage{times}
\usepackage{microtype}

\usepackage{amsmath,amssymb,mathtools}
\usepackage{algorithm}
\usepackage{algpseudocode}
\usepackage{graphicx}
\usepackage[caption=false,font=footnotesize]{subfig}
\usepackage{booktabs,multirow,tabularx,array,makecell,threeparttable}
\usepackage{capt-of}
\usepackage{placeins}
\usepackage{siunitx}
\usepackage{xcolor}

\usepackage{cite}
\usepackage{balance}
\usepackage[hidelinks]{hyperref}

\newcommand{\FigureImage}[2]{%
  \parbox[c][\dimexpr#1+2\fboxsep+2\fboxrule\relax][c]{\linewidth}{%
    \centering
    \includegraphics[
      width=\dimexpr\linewidth-2\fboxsep-2\fboxrule\relax,
      height=#1,
      keepaspectratio
    ]{#2}%
  }%
}

\title{KORD: Breaking the Key-Generation Bottleneck in Dealerless FSS via Protocol--Hardware Co-Design}

\author{\normalfont\normalsize
  \begin{tabular}{c}
	Yijing Peng$^{\dagger\ddagger}$, Lin Liu$^{\dagger\ddagger*}$, Yujie Xue$^{\dagger}$, Shaojing Fu$^{\dagger}$, Shaoqing Li$^{\dagger\ddagger}$, Yaohua Wang$^{\dagger\ddagger}$, Rongmao Chen$^{\dagger}$, Yang Guo$^{\dagger\ddagger}$\\[4pt]
	$\dagger$~College of Computer Science and Technology, National University of Defense Technology, Changsha, China\\
	$\ddagger$~Key Laboratory of Advanced Microprocessor Chips and Systems, National University of Defense Technology, Changsha, China
\end{tabular}
\thanks{$^{*}$~Corresponding author: Liu Lin (liulin16@nudt.edu.cn)}
}

\begin{document}
\pagestyle{plain}
\maketitle
\thispagestyle{plain}
\begin{abstract}
\bfseries
Function secret sharing (FSS) has become a core primitive in privacy-preserving computation. However, each FSS invocation requires a fresh pair of function keys generated by a trusted dealer , expands the system's trust boundary and hinders practical deployment. Existing dealerless protocols eliminate this dependency, but incur substantial communication and a number of interaction rounds that grows linearly with the input bit-width, making key generation a major bottleneck.

This paper present KORD, a protocol--hardware co-design that dramatically reduces the cost of dealerless FSS key generation. At its core is a pair of special-purpose chips that establish a common root of trust through mutual attestation and, within it, reconstruct FSS keys---eliminating the need for a dealer. This root of trust further forms a security boundary within which KORD restructures the generation protocol, collapsing the interaction of prior dealerless protocols into a single round, independent of GGM depth. A cross-key scheduling scheme then interleaves independent GGM-tree traversals, sustaining high computational throughput. KORD reduces per-key-generation communication by 7,633--70,274$\times$ over the state-of-the-art distributed FSS protocol across a comprehensive suite of FSS building blocks. Post-route analysis projects 12.75 million 32-bit DPF keys per second at 204 MHz using 21.5K LUTs, with 99.8\% AES lane utilization. On private ResNet-18 inference, KORD cuts the share of end-to-end time spent on key generation from over 96\% to 11.9\%.
\end{abstract}

\section{Introduction}
\label{sec:introduction}

Function Secret Sharing (FSS) has recently become an increasingly prevalent and powerful privacy-preserving computing technique, which has been extensively adopted in secure two-party computation (2PC) and various real-world privacy-sensitive scenarios \cite{ref21,ref25,ref48} due to its outstanding online efficiency and low communication overhead \cite{ref7,ref8,ref17}. Similar to mainstream preprocessing-based 2PC frameworks, FSS-based secure computation decouples the entire computing process into two independent stages, i.e, an offline preprocessing stage and an online evaluation stage. The offline stage performs all heavy cryptographic computation to generate input-independent correlated randomness, i.e., a key pair. The production of which usually relies on a trusted dealer. The online stage only executes lightweight computing operations based on the pre-generated randomness, effectively reducing online interaction latency and communication pressure.

Despite the remarkable efficiency advantages, the dealer-based conventional FSS-based 2PC protocols weaken the system's security guarantees. The secure generation and synchronization of such paired keys become the core bottleneck of the whole system, and the inherent dealer dependency also brings unavoidable trust risks, single-point failures and deployment restrictions, which greatly limit the practical applicability and security of FSS privacy-preserving systems.

To tackle such problems, researchers also propose dealer-free deployment schemes \cite{ref15,ref20,ref54}. Floram \cite{ref15} pioneers a work where the FSS key could be generated without a dealer, then proposes the first dealer-less Distribution Point Function (DPF). After that, Half-tree further optimizes the DPF and then present a dealer-less Distribution Comparison Function (DCF). Unfortunately, neither of them can support arithmetic-shared inputs or outputs which limits their applications, such as privacy-preserving machine learning inference or training \cite{ref14,ref21}.

Most recently, Xing et al. \cite{ref54} proposed a 2PC FSS scheme which can generate FSS keys for DPF and DCF without a dealer, supporting both arithmetic-shared inputs and outputs. The key limits of this work lie in two aspects:
\begin{itemize}
  \item \textbf{Linearly growing communication.} At the heart of every FSS key is a GGM tree \cite{ref18}, a binary tree of pseudorandom seeds expanded one level per bit of the $\ell$-bit input, where each level publishes a small correction word computed jointly from both parties' secret seeds. A dealerless protocol must therefore run an interactive at every level, so communication and interaction rounds grow linearly with $\ell$, which its own evaluation places at seconds per key over a WAN \cite{ref54}.
  \item \textbf{Full-tree expansion.} Since the hidden point is secret-shared, neither party knows which path of the tree is the special one, so the Gen algorithm must expand the full GGM tree of $2^\ell$ leaves for every key (Algorithm~2 of \cite{ref54}); generation computation therefore grows exponentially with the input width, even when communication is discounted.
\end{itemize}

Both costs recur for every one of the millions of keys an application consumes, making key generation the major bottleneck of dealerless FSS deployment.

To address these challenges, we propose KORD, a protocol--hardware co-design that confines the joint key-generation state of the 2PC to a pair of special-purpose chips, which establish a common root of trust during a one-time setup. Moreover, each chip reconstructs the key's parameters inside its security boundary and expands only the special path, reducing per-key computation from exponential to linear in $\ell$. Since different keys have independent seed chains, KORD introduces a cross-key scheduling fabric that interleaves many in-flight keys, filling one key's dependency bubbles with another's ready work and raising lane utilization from 8.3\% to 99.8\% in cycle-accurate RTL simulation.

The contributions of this work can be summarized as follows.
\begin{itemize}
  \item We propose KORD, a new protocol--hardware co-design that generates FSS keys without any dealer, reducing per-key communication to a fixed single round, independent of the input length, while cutting per-key computation from exponential full-tree expansion to a linear walk of the special path.
  \item We introduce a cross-key scheduling design that interleaves independent GGM traversals through a single shared AES lane array, raising lane utilization from 8.3\% to 99.8\% in cycle-accurate RTL simulation, a saturation behavior confirmed on silicon.
  \item We implement KORD end-to-end and validate it on a physical ZCU102 board, where 60 keys reproduce the golden stream bit-exactly, every fail-closed attack class is rejected, and the online path runs bit-exact past 214~MHz. Across nine FSS building blocks, KORD reduces per-operation key-generation communication by $7{,}633$--$70{,}274\times$ over the state-of-the-art dealerless protocol \cite{ref54}; the synthesized fabric projects 12.75 million DPF-32 keys/s; and on private ResNet-18 inference, the key-generation share of end-to-end time falls from over 96\% to 11.9\%.
\end{itemize}

\section{Background and Threat Model}
\label{sec:background-threat-model}

\subsection{FSS, DPF/DCF, and the GGM Construction}
\label{subsec:fss-background}

Table~\ref{tab:notation} summarizes the notation used throughout the paper. A two-party FSS scheme \cite{ref8,ref17} for a function family $\mathcal{F}=\{f:\{0,1\}^{\ell}\rightarrow\mathbb{G}\}$ over an Abelian group $\mathbb{G}$ consists of two algorithms, $(\mathsf{Gen},\mathsf{Eval})$. The key-generation algorithm $\mathsf{Gen}(1^\lambda,\hat{f})$ produces a key pair $(k_0,k_1)$, while $\mathsf{Eval}(b,k_b,x)$ denotes the local evaluation performed by party $P_b$. Correctness requires
\begin{equation}
  \mathsf{Eval}(0,k_0,x)+\mathsf{Eval}(1,k_1,x)=f(x)
  \quad\text{for all }x\in\{0,1\}^{\ell}.
  \label{eq:fss-correctness}
\end{equation}
Security requires the distribution of either key in isolation to be simulatable given only the public parameters $(\lambda,\ell,\mathbb{G})$.

Distributed point functions (DPFs) \cite{ref8,ref17} and distributed comparison functions (DCFs) \cite{ref4,ref8} instantiate FSS for point and comparison functions, respectively:
\begin{equation}
  f_{\alpha,\beta}(x)=
  \begin{cases}
    \beta, & x=\alpha,\\
    0, & \text{otherwise},
  \end{cases}
  \qquad
  f^{<}_{\alpha,\beta}(x)=
  \begin{cases}
    \beta, & x<\alpha,\\
    0, & \text{otherwise}.
  \end{cases}
  \label{eq:dpf-dcf-functions}
\end{equation}
Here, $\alpha\in\{0,1\}^{\ell}$ is the hidden point and $\beta\in\mathbb{G}$ is the payload; throughout this paper, $\mathbb{G}=\mathbb{Z}_{2^{\ell_{\mathrm{out}}}}$. Equality and point lookup reduce to DPF evaluation, whereas comparison, ReLU, and truncation reduce to DCF evaluation \cite{ref4,ref22}.

\begin{table}[t]
  \caption{Notation.}
  \label{tab:notation}
  \centering
  \footnotesize
  \setlength{\tabcolsep}{3pt}
  \begin{tabular}{@{}p{0.27\columnwidth}p{0.66\columnwidth}@{}}
    \hline
    Symbol & Meaning \\
    \hline
    $\lambda$ & security parameter (128 throughout) \\
    $\ell,\ell_{\mathrm{out}}$ & input and output bit width of a gate \\
    $\alpha,\beta$ & hidden point and payload of a key \\
    $P_0,P_1;\ C_0,C_1$ & the two parties and their KORD chips \\
    $\langle x\rangle_b$ & $P_b$'s additive share of $x$ \\
    $k_b$ & $P_b$'s FSS key share \\
    $G$ & length-doubling PRG (fixed-key AES) \\
    $s_b^j,t_b^j$ & level-$j$ seed and control bit of $P_b$ \\
    $\mathit{CW}_j,\mathit{CW}^{V}_j$ & correction word and value correction word \\
    $K_{\mathrm{ch}}$ & shared root seed (master, set at pairing) \\
    $k_r,k_p$ & $K_{\mathrm{ch}}$ subkeys: root derivation; KORD-mac channel \\
    $K_{\mathrm{otp}}$ & session keystream secret (default channel) \\
    $\mathit{ctr}$ & monotonic message / per-key counter \\
    $W,M$ & fabric lanes and in-flight key slots \\
    \hline
  \end{tabular}
\end{table}

Both primitives use a binary GGM tree \cite{ref18} of depth $\ell$. A length-doubling PRG, $G(s)=(s^L,t^L,s^R,t^R)$, advances each party by one level for every input bit and is typically instantiated with fixed-key AES \cite{ref37}. Traversal begins with an independent root seed $s_b^0$ and control bit $t_b^0=b$ for each party. At level $j$, the generator expands both parties' level-$(j-1)$ seeds and publishes the correction word
\begin{equation}
  \mathit{CW}_j=\left(
    s_0^{\overline{\alpha_j}}\oplus s_1^{\overline{\alpha_j}},
    t_0^L\oplus t_1^L\oplus\alpha_j\oplus 1,
    t_0^R\oplus t_1^R\oplus\alpha_j
  \right),
  \label{eq:seed-correction-word}
\end{equation}
where $\alpha_j$ is the $j$-th bit of $\alpha$ and $\overline{\alpha_j}$ selects the child off the evaluation path. Each party applies $\mathit{CW}_j$ exactly when its control bit is set. Off the path, the parties' seeds and control bits coincide and cancel in Eq.~\eqref{eq:fss-correctness}; on the path, they diverge. The final correction word maps this divergence to shares of the payload:
\begin{equation}
  \mathit{CW}_{\ell+1}=(-1)^{t_1^\ell}
  \left(\beta-\mathsf{Convert}(s_0^\ell)+\mathsf{Convert}(s_1^\ell)\right),
  \label{eq:final-correction-word}
\end{equation}
where $\mathsf{Convert}$ is a pseudorandom map from seeds to $\mathbb{G}$. A DCF key additionally includes one value correction word $\mathit{CW}^{V}_j$ per level, derived from the same joint seed state \cite{ref8}; its correctness lemma appears in the anonymous supplementary material.

A DPF key share occupies $\lambda+\ell(\lambda+2)+\ell_{\mathrm{out}}$ bits, while a DCF key share adds one $\ell_{\mathrm{out}}$-bit value correction word per level \cite{ref4,ref8}. Two structural properties determine key-generation cost. \textbf{(P1) Depth-serial expansion:} level $j$ consumes the seed produced at level $j-1$, making each key a serial chain of $\ell$ dependent AES calls and creating a hardware latency wall. \textbf{(P2) One-time keys:} $\alpha$ is bound to the fresh mask of the value consumed by the gate; reusing one key for two inputs therefore breaks hiding. Keys are consumable objects, so any key source must sustain the workload's full gate rate.

In the online phase, each gate requires one local evaluation and one round to open a masked value using one or two ring elements \cite{ref8,ref9,ref22}. All remaining costs arise during key generation. In a dealerless design, key generation also incurs the cryptographic interaction required to construct Eq.~\eqref{eq:seed-correction-word} from both parties' seed chains at every level.

\subsection{Threat Model and Security Boundary}
\label{subsec:threat-model}

\textbf{Parties and corruption model.} $P_0$ and $P_1$ execute a two-party computation under the semi-honest model, which is also assumed by the dealerless baseline. All analyses and reported results use this model. Malicious parties are outside our scope; extending KORD to this setting, for example by batching MACs over key requests, remains future work.

\textbf{System components.} Each party's machine contains one special-purpose KORD chip: $C_0$ at $P_0$ and $C_1$ at $P_1$. The chips load no executable code and accept only the protocol's fixed message format. Neither chip has a network interface; instead, the host stacks relay an end-to-end logical channel whose messages are processed by the secure-channel units.

\textbf{Adversary capabilities.} We consider three adversarial positions. First, a semi-honest party observes its own inputs and shares, the key shares emitted by its chip, and the protocol transcript; the simulation-based analysis requires this view to reveal nothing further. Second, the inter-site network, including all relay software, is semi-honest in the default deployment, matching the model of the parties and of the dealerless baseline: it may read every message but delivers faithfully. Confidentiality against this path rests on the one-time session keystream; integrity is an assumption rather than an enforcement, since a keystream mask is malleable. The hardened KORD-mac deployment removes exactly this assumption: the network may drop, modify, reorder, or replay messages, the secure channel rather than the transport provides integrity and replay protection, and a network adversary cannot cause a chip to accept a forged, modified, replayed, or cross-substituted frame (Section~VI). Each host handles its own party's plaintext and, like the party itself, is assumed semi-honest. Consequently, share modification by a host constitutes a protocol deviation and is outside the model. Third, a physical adversary that opens either chip defeats the scheme. Pairing-time attestation prevents device substitution at deployment but does not protect against invasive attacks during the device's lifetime.

\textbf{Trust anchors.} A one-time attested pairing at deployment establishes the system secrets. The chips mutually attest \cite{ref13,ref24} and agree on a shared root seed $K_{\mathrm{ch}}$, the session keystream secret $K_{\mathrm{otp}}$ that encrypts every inter-site message in the default channel, and initial monotonic counters $\mathit{ctr}$. Thereafter, both chips apply the same domain-separated key schedule to derive identical material from $K_{\mathrm{ch}}$ without further communication. The channel subkey $k_p$ encrypts and authenticates every inter-site message using AES-GCM in the hardened KORD-mac deployment. The root subkey $k_r$, combined with a per-key counter advanced in lockstep, derives the two DPF/DCF root seeds $s_0^0$ and $s_1^0$. Domain separation keeps $k_p$ and $k_r$ computationally independent. Root expansion uses the fixed, public AES-MMO convention; FSS security relies on the secrecy of the roots rather than the expander. After pairing, the hosts observe only protocol messages. We model attestation as an imported provisioning primitive based on symmetric pre-shared master keys. Certificate chains, measured boot, and physical entropy are assumptions rather than mechanisms established by KORD; accordingly, the model provides attestable pairing rather than a complete certificate-backed attestation protocol.

\textbf{Security boundary.} The security boundary comprises the two chips and the logical channel connecting them; the parties, hosts, and physical network remain outside. Plaintext copies of $(\alpha,\beta)$, the seed chains used in Eq.~\eqref{eq:seed-correction-word}, and the root seed exist only within this boundary. All inter-site traffic is authenticated ciphertext, and each party exchanges only its own input shares and resulting key share with its local chip. Each chip is trusted to perform exactly three functions: authenticate and decrypt requests; reconstruct $(\alpha,\beta)$ and expand it into standard key shares; and emit those shares while zeroizing all intermediate state. The minimality argument treats each of these functions as necessary.

\textbf{Cross-party trust.} $P_0$ must trust the chip deployed at $P_1$'s site, and $P_1$ must symmetrically trust the chip at $P_0$'s site. Because each chip reconstructs the complete $(\alpha,\beta)$ and holds $K_{\mathrm{ch}}$, compromising either chip exposes both parties' keys and masks associated with that pairing. This exposure is the inherent cost of the minimal-hardware-root approach. Under our model, the trusted component is vulnerable only to physical compromise, whereas the dealer it replaces is also exposed online. A party that cannot audit or attest the remote chip has no recourse within this model.

\section{Motivation}
\label{sec:motivation}

\subsection{FSS Requires an Application-Rate Key Stream}
\label{subsec:application-rate-key-stream}

FSS shifts online nonlinear computation into the consumption of one-time key material. Each comparison, lookup, ReLU, truncation, or oblivious access consumes one or more fresh key pairs, and independent operations cannot reuse them. Large private-inference and query workloads therefore create a sustained key stream rather than a finite startup batch. In the dealerless ResNet-18 workload evaluated later, key generation accounts for at least 96.9\% of end-to-end time; the full workload model and baseline methodology appear in Section~VII-C.

Correctness alone is consequently insufficient for a deployable key source. It must satisfy three system requirements: (1) no dealer; (2) a number of per-key interaction rounds independent of the input bit-width $\ell$; and (3) a key-supply rate that matches application demand. The first two govern deployment and network scalability, while the third determines whether key generation remains off the application's critical path.

\subsection{Existing Key Sources Trade Trust for Communication}
\label{subsec:key-source-tradeoff}

\begin{figure*}[t]
  \begin{minipage}[t]{0.485\textwidth}
    \centering
    \FigureImage{2.15in}{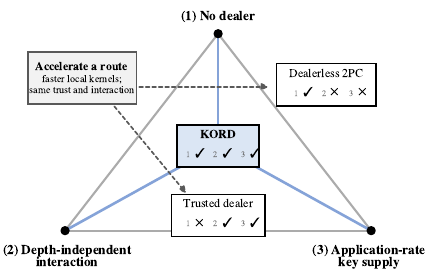}
    \captionof{figure}{Design space for FSS key sources. Dealer-based and dealerless-2PC routes trade among no dealer, depth-independent interaction, and application-rate key supply. Accelerating a route does not change its trust or interaction structure. KORD targets all three under a minimal hardware root. The trilemma is descriptive, not an impossibility result.}
    \label{fig:design-space}
  \end{minipage}\hfill
  \begin{minipage}[t]{0.485\textwidth}
    \centering
    \FigureImage{2.85in}{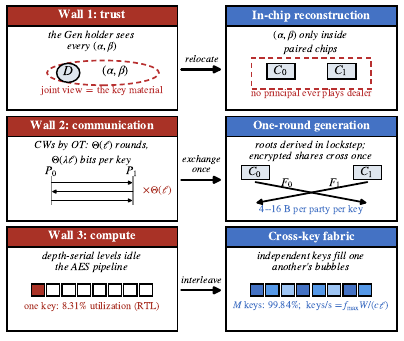}
    \captionof{figure}{KORD design overview. Each wall identified in this section (left) maps to the KORD mechanism that removes it (right): relocate the joint view into paired chips, exchange once over the network, and interleave independent keys through a shared array (Sections~IV and~V).}
    \label{fig:kord-overview}
  \end{minipage}
\end{figure*}

Deployed FSS systems follow three routes (Figure~\ref{fig:design-space}). A trusted dealer \cite{ref22} samples both keys centrally, enabling low interaction and centralized acceleration. The dealer, however, observes the joint key-generation state and becomes an additional trusted principal and single point of compromise. Dealerless 2PC \cite{ref15,ref54} removes that principal by computing GGM correction words jointly, but existing protocols emulate this computation level by level. Their interaction rounds therefore grow with GGM depth, and their communication grows as $\Theta(\lambda\ell)$ per key, limiting the attainable key-supply rate over a network.

A third route accelerates the local kernels of either design with CPUs, GPUs, or AES datapaths \cite{ref21,ref25,ref53}. Existing acceleration improves the execution of a route, but does not change its trust and interaction structure: it changes neither who holds the joint key-generation state nor whether the parties interact at every level. The cost is substantial in practice. At $\ell=16$, our reproduction of the state-of-the-art dealerless DPF generator \cite{ref54} transfers roughly 80~KB per party across hundreds of messages; the complete formulas, reproduction procedure, and extrapolation policy appear in Section~VII-A. These routes define the descriptive trilemma in Figure~\ref{fig:design-space}; they do not establish a formal impossibility result.

\subsection{Structural Causes and Architectural Implications}
\label{subsec:structural-causes}

The tradeoff above has a structural cause. Resolving it moves the bottleneck from communication into computation and yields three requirements for the architecture (Figure~\ref{fig:kord-overview}).

\textbf{Observation 1: the joint view couples trust and communication.} A DPF/DCF generator forms each correction word from the joint view of both parties' seed chains. This joint key-generation state is sensitive because it determines the complete function. Giving it to an independent principal creates a dealer; emulating access to it between the parties through existing 2PC constructions creates per-level interaction. Faster OT, AES, or PRG implementations reduce the cost of each step but do not change where the joint state resides or how many levels require interaction. A minimal protected, special-purpose boundary offers a third location for this state. \textbf{Design Requirement 1:} joint key-generation state must remain inside a minimal protected boundary. \textbf{Design Requirement 2:} the number of per-key network interaction rounds must be independent of the GGM depth $\ell$.

\textbf{Observation 2: relocating generation exposes a compute wall.} Confining the joint view to hardware removes the level-by-level network emulation, but the protected boundary then inherits a depth-serial GGM traversal. Level $j$ expands the seed produced at level $j-1$, so an isolated key cannot issue its next level until the AES pipeline returns the previous result. In cycle-accurate RTL simulation, an isolated key occupies KORD's pipelined AES lanes in only 8.31\% of issue cycles. Replicating complete engines can increase throughput, but scales area and state storage with the number of replicas.

\textbf{Observation 3: independent keys expose cross-key parallelism.} The dependence is unavoidable within one key, but different keys have independent seed chains. Interleaving multiple GGM traversals through a shared array can fill one key's dependency bubbles with another key's ready work. The architecture should therefore optimize sustained key-supply throughput rather than only single-key latency. \textbf{Design Requirement 3:} the architecture must exploit cross-key parallelism to sustain the application's key-consumption rate.

These observations yield three requirements: confine the joint view to a minimal protected boundary, make per-key interaction independent of GGM depth, and exploit cross-key parallelism to overcome serial GGM dependence. The following architecture realizes these requirements through protocol--hardware co-design.

\section{The KORD Architecture and Core Design}
\label{sec:kord-architecture}

\subsection{Key Design Goals}
\label{subsec:key-design-goals}

KORD must generate FSS keys \emph{without a dealer}. At no phase may any single machine hold the point--payload pair $(\alpha,\beta)$ or both keys. KORD must also emit keys bit-identical to a dealer's so that evaluation and every higher software layer run unchanged. Its goals map one-to-one to the three walls in Figure~2; Section~\ref{sec:hardware-architecture} implements them in hardware:

\begin{itemize}
  \item \emph{In-chip reconstruction} (wall 1, trust). The parties hold only additive shares of $(\alpha,\beta)$; the pair may be reconstructed only where neither party, nor any other machine, can observe it, so no principal ever plays dealer.
  \item \emph{One-round generation} (wall 2, communication). A key must cost a single exchange of fixed-size encrypted messages, with no dependence of the round count on the bit width $\ell$.
  \item \emph{Cross-key supply} (wall 3, compute). The chip inherits the dealer's depth-serial GGM expansion; by interleaving independent keys through one array it must still supply keys at the rate the online phase consumes them.
\end{itemize}

\begin{algorithm}[t]
  \caption{KORD one-round DPF key generation (semi-honest); \textcolor{red}{red} marks the single network round.}
  \label{alg:kord-dpf-generation}
  \begin{algorithmic}[1]
    \Require each party $P_b$ holds $\langle\alpha\rangle_b,\langle\beta\rangle_b$; the paired chips hold the session keystream secret $K_{\mathrm{otp}}$, the $K_{\mathrm{ch}}$-derived subkeys $k_p,k_r$, and matching counters $\mathit{ctr}$
    \Ensure each $P_b$ obtains its standard DPF key share $k_b$
    \State $P_b$ (host): write the share pair $(\langle\alpha\rangle_b,\langle\beta\rangle_b)$ into the local chip $C_b$; a party's own input crosses only its own local interface, never the network in plaintext
    \State $C_b$: encrypt $F_b\gets(\langle\alpha\rangle_b\mathbin\Vert\langle\beta\rangle_b)\oplus\mathsf{KS}(K_{\mathrm{otp}},\mathit{ctr}_b)$ with the one-time session keystream (KORD-mac: $F_b\gets\mathsf{AE}_{k_p}(\cdot)$ with mode, widths, and $\mathit{ctr}_b$ bound as authenticated header fields); return the ciphertext $F_b$ to the host; the channel secrets never leave the chips, so no host can open a frame
    \State \textcolor{red}{$P_0\leftrightarrow P_1$ (the single round): $P_0$ sends $F_0$ to $P_1$ while $P_1$ sends $F_1$ to $P_0$; only ciphertext crosses the network}
    \State $P_b$ (host): forward the received peer frame into $C_b$, so each chip holds both $F_0$ and $F_1$
    \State $C_b$: bind each frame to the lockstep counter $\mathit{ctr}_b$ (KORD-mac: verify both tags under $k_p$ and both transported counters; any failure aborts fail-closed, with no decryption and no engine dispatch)
    \State $C_b$: decrypt both frames; reconstruct $\alpha\gets\langle\alpha\rangle_0+\langle\alpha\rangle_1$, $\beta\gets\langle\beta\rangle_0+\langle\beta\rangle_1$; the pair now exists, and only here
    \State $C_b$: derive the roots $(s_0^0,s_1^0)\gets\mathsf{KDF}(k_r,\mathit{ctr})$, identical in both chips by lockstep, so the roots are never transmitted; $t_0^0\gets0$, $t_1^0\gets1$
    \For{$j=1$ \textbf{to} $\ell$}
      \State expand $G(s_0^{j-1})$ and $G(s_1^{j-1})$ with the fixed public PRG; form $\mathit{CW}_j$ by Eq.~(3)
      \State advance both seed chains along the path to $\alpha$
    \EndFor
    \State compute $\mathit{CW}_{\ell+1}$ by Eq.~(4); the loop is the algorithm's only $\Theta(\ell)$ stage, depth-serial within a key; independent keys interleave through it in hardware
    \State $C_b$: $\mathit{ctr}\gets\mathit{ctr}+1$; assemble $k_b\gets(s_b^0,\mathit{CW}_1,\ldots,\mathit{CW}_{\ell+1})$; zeroize $(\alpha,\beta)$ and all seeds
    \State \Return $k_b$ to $P_b$ only, over the local interface
  \end{algorithmic}
\end{algorithm}

\begin{algorithm}[t]
  \caption{KORD one-round DCF key generation (semi-honest); step 1 folds in the shared admission of Algorithm~\ref{alg:kord-dpf-generation}.}
  \label{alg:kord-dcf-generation}
  \begin{algorithmic}[1]
    \Require as in Algorithm~\ref{alg:kord-dpf-generation}, with the DCF mode and widths bound in the authenticated header fields
    \Ensure each $P_b$ obtains its standard DCF key share $k_b$
    \State run steps 1--6 of Algorithm~\ref{alg:kord-dpf-generation} unchanged: submit shares, encrypt $F_b$, \textcolor{red}{exchange in the single round}, forward, bind counters (KORD-mac: verify fail-closed), decrypt and reconstruct $(\alpha,\beta)$
    \State $C_b$: derive the roots $(s_0^0,s_1^0)\gets\mathsf{KDF}(k_r,\mathit{ctr})$; $t_0^0\gets0$, $t_1^0\gets1$; $V_\alpha\gets0$
    \For{$j=1$ \textbf{to} $\ell$}
      \State expand $(s_{b'}^L,v_{b'}^L,s_{b'}^R,v_{b'}^R)\gets G'(s_{b'}^{j-1})$ for both chains $b'\in\{0,1\}$, four PRG calls per chain, the value lanes $v^L,v^R$ feeding the per-level value corrections; Keep/Lose from bit $\alpha_j$
      \State form the seed correction $\mathit{CW}_j$ from the Lose-side seeds by Eq.~(3)
      \State form the value correction
      \Statex \hspace{1em}$\displaystyle
        \begin{aligned}
          \mathit{CW}_j^V &\gets (-1)^{t_1^{j-1}}\bigl[
            \operatorname{convert}(v_1^{\mathrm{Lose}})
            -\operatorname{convert}(v_0^{\mathrm{Lose}})\\
          &\hspace{5.6em}{}-V_\alpha
            +\mathbf{1}_{[\mathrm{Lose}=L]}\beta\bigr];
        \end{aligned}$
      \Statex \hspace{1em}update the running $V_\alpha$ by the recurrence of \cite{ref4}
      \State advance both seed chains along the path to $\alpha$
    \EndFor
    \State compute the final correction $\mathit{CW}_{\ell+1}\gets(-1)^{t_1^\ell}\left[\operatorname{convert}(s_1^\ell)-\operatorname{convert}(s_0^\ell)-V_\alpha\right]$
    \State $C_b$: $\mathit{ctr}\gets\mathit{ctr}+1$; assemble $k_b\gets(s_b^0,\{\mathit{CW}_j,\mathit{CW}_j^V\}_{j=1}^{\ell},\mathit{CW}_{\ell+1})$; zeroize $(\alpha,\beta)$, $V_\alpha$, and all seeds
    \State \Return $k_b$ to $P_b$ only, over the local interface
  \end{algorithmic}
\end{algorithm}

\subsection{One-Round Key Generation}
\label{subsec:one-round-generation}

A DPF/DCF key pair is a deterministic function of three inputs: the reconstructed $(\alpha,\beta)$, the two roots $s_0^0,s_1^0$, and the PRG. Once paired, both chips derive identical roots from $K_{\mathrm{ch}}$ through $k_r$ and a per-key counter. They also share the fixed public AES-MMO convention. The share pair is therefore the only per-key input that a chip lacks, and it must come from both parties. One simultaneous exchange of the two share messages is both necessary and sufficient. After the exchange, each chip holds bit-identical inputs, independently computes the same correction-word stream, and emits only its own party's key (Figure~\ref{fig:one-round-exchange}). This design eliminates the $\Theta(\ell)$ rounds of oblivious-transfer interaction. Each chip performs the same expansion as a dealer, so its cost matches the dealer's.

\emph{Prerequisite: attested pairing.} Generation assumes that the two chips already share $K_{\mathrm{ch}}$. KORD establishes this secret once, outside the per-key path, using the secure-initialization protocol of PPMLAC \cite{ref56} unchanged. Domain-separated KDFs then yield $k_r$, $k_p$, and zeroed counters. Both sides of that protocol contribute entropy, so a host can neither bias nor replay the seed, and no host ever holds any derived secret.

\begin{figure}[t]
  \centering
  \FigureImage{2.80in}{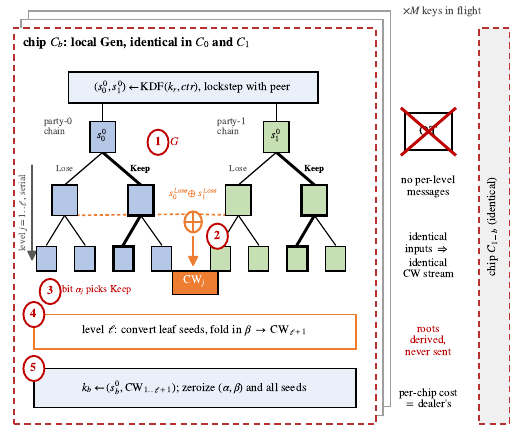}
  \caption{After one exchange, each chip runs the identical local Gen over \emph{both} parties' seed chains (4-leaf tree shown). The chips derive the roots in lockstep from $(k_r,\mathit{ctr})$ and never transmit them. \textcircled{1} Expand each level with the public AES-MMO $G$. \textcircled{2} XOR the two Lose-side seeds into $\mathit{CW}_j$ (Eq.~(3)). \textcircled{3} Advance along the path to $\alpha$; its bit selects the Keep child. \textcircled{4} Convert the leaf seeds into the final correction (Eq.~(4)). \textcircled{5} Assemble and return only the party's own key. Levels are depth-serial within a key. No oblivious transfer or per-level message crosses between the chips.}
  \label{fig:local-generation}
\end{figure}

\begin{figure}[t]
  \centering
  \FigureImage{2.05in}{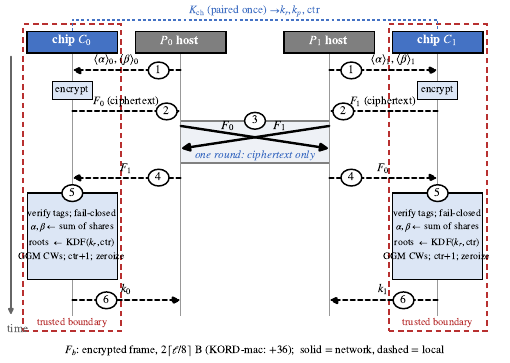}
  \caption{The one-round exchange in time order. \textcircled{1} Each party submits its own share pair to its own chip over the local interface. \textcircled{2} The chip returns the encrypted frame $F_b$; only this ciphertext ever leaves the trust boundary. \textcircled{3} The hosts exchange $F_0,F_1$, the protocol's only network round. \textcircled{4} Each host forwards the peer's frame back into its own chip. \textcircled{5} Both chips verify, reconstruct, and expand alone. \textcircled{6} Each chip emits its own party's key. Solid arrows cross the network; dashed arrows are local host--chip transfers. Red dashed lines mark the trust boundaries.}
  \label{fig:one-round-exchange}
\end{figure}

\emph{The exchange.} The parties hold the bind point and payload as additive shares, $\alpha=\alpha_0+\alpha_1$ and $\beta=\beta_0+\beta_1$. Neither party knows $\alpha$ or $\beta$. Each party $P_b$ submits its share pair to its own chip over the local interface, exposing only its own input. The chip encrypts the pair into one ciphertext frame that is exactly the two share fields, $F_b=(\langle\alpha\rangle_b\mathbin\Vert\langle\beta\rangle_b)\oplus\mathsf{KS}(K_{\mathrm{otp}},\mathit{ctr}_b)$: $2\lceil\ell/8\rceil$ bytes, or 4/8/16~B at $\ell=16/32/64$, with no per-frame nonce or tag. The FSS mode and widths are public session parameters fixed at pairing, and freshness comes from the counters both chips advance in lockstep; we set $\ell_{\mathrm{out}}=\ell$ throughout, and aggregate network traffic per key is the two frames together. The hardened KORD-mac deployment instead wraps the same fields in per-message authenticated encryption under $k_p$, $[\text{nonce }12\mid\mathit{ctr}\ 8\mid\text{ct }2\lceil\ell/8\rceil\mid\text{tag }16]$, a fixed 36-byte envelope (40/44/52~B) with mode and widths as associated data; under it the SCU authenticates both messages before any further processing, fail-closed: no decryption and no engine dispatch (Section~\ref{subsec:realizing-security-boundary}). The hosts exchange the two messages in one simultaneous round, after which each chip holds $\{F_0,F_1\}$. Nothing else crosses the network for that key. Algorithms~\ref{alg:kord-dpf-generation} and~\ref{alg:kord-dcf-generation} give the DPF and DCF bodies both chips then execute. In both algorithms, every step after the single exchange is local, the expansion loop is the unchanged dealer core \cite{ref8}, and the DCF value corrections follow \cite{ref4}. Each chip emits its own party's key share $k_b$ over the local interface, which is the party's own protocol output, and advances its per-key counter. The counters remain matched as long as both chips accept the same message sequence. A network that makes the sequences diverge yields unusable keys, not leaked ones.

\emph{From the algorithms to the hardware.} The one-round property and the hardware boundary coincide. Steps 1, 3, and 4 (submit, exchange, forward) constitute the software's entire role. Software holds no secret beyond the party's own shares and emitted key share; every remaining step touches a secret and executes in hardware. The mapping onto the datapath is one-to-one, using the step numbers in Algorithm~\ref{alg:kord-dpf-generation}. Steps 2 and 5--6 map to the SCU and reconstruction zone, step 7 to the root-derivation unit, the loop in steps 8--12 to the fabric, and step 13 to the assembler and counter commit (Section~\ref{subsec:datapath-dataflow}). The host may batch submissions and keep many keys in flight because each message is bound to its lockstep counter (under KORD-mac the counter is authenticated associated data, and the chip rejects misordered frames fail-closed). Within the hardware, the loop's only data dependency is confined to one key. The fabric converts this property into throughput.

\emph{Evaluation.} The emitted keys are standard DPF/DCF key shares in the ordinary BGI-style layout \cite{ref8}. They are bit-exact and interchangeable with dealer-generated keys, so the existing online stack runs unchanged on KORD with the same binaries and key semantics (verified in Section~VII-C). Each party evaluates locally.

The PRG is instantiated as fixed-key AES-128 in Matyas--Meyer--Oseas (MMO) mode \cite{ref2}, the standard choice for hardware FSS pipelines \cite{ref1,ref19,ref25}. This PRG is functionally equivalent to, but not bit-compatible with, software stacks that default to per-node re-keyed AES. Interoperability therefore fixes the convention at pairing time. We validated the convention end-to-end. The RTL, our reference software, and the LLAMA stack under the MMO switch produce identical keys (101,334 field checks, zero mismatches). KORD-generated keys also drive a private ResNet-18 inference whose 1,000 unmasked logits exactly match the validated reference. This is a convention check under rebuilt binaries, separate from the end-to-end deployment run.

\subsection{Design Rationale}
\label{subsec:design-rationale}

\emph{Transmitting bound shares.} KORD's default deployment transmits $(\alpha,\beta)$ shares even though some protocols allow one party to choose the mask under an FSS key. A self-chosen-mask deployment would reduce even KORD's residual communication to nearly zero. Three properties motivate the default. \emph{Generality:} in our workloads the bound point is a joint secret existing only as two shares (a DORAM index, a compiled program's wire mask). No party may know $\alpha$. The shares must be combined, and the reconstruction zone is the one place where this can occur without re-creating a dealer. \emph{Compatibility:} the chip consumes true shares and emits standard key shares, making KORD a drop-in key source. Self-chosen masks instead change the protocol semantics around every FSS gate. \emph{Model-aligned accounting:} the default channel assumes exactly the semi-honest setting of the parties and of the baseline, so the headline comparison is like-for-like; the hardened KORD-mac tier prices the malicious-network upgrade at a fixed 36 bytes per message. Self-chosen-mask protocols can reduce key-generation communication under KORD to essentially zero.

\emph{Shared-root-seed derivation.} The one-round property depends on never transmitting the per-key roots. This design is sound for two reasons. (i) The root seed is created at attested pairing and resides only inside the two trust boundaries, so the derivation has an empty network view. (ii) Both chips run the same deterministic derivation, making their roots identical by construction. Indistinguishability from ideal randomness then reduces to the pseudorandomness of the derivation function, the assumption class on which FSS security rests. The corresponding simulation step is the chip segment of the composition in the supplementary material.

\emph{The KORD-mac deployment.} KORD's default channel rests on the two assumptions stated in Section~II-B: a semi-honest network path and a one-shot session keystream. KORD-mac drops both. It wraps every message in per-message authenticated encryption under the pairing-derived subkey $k_p$, a fixed 36-byte envelope per frame (40/44/52~B at $\ell=16/32/64$), and in exchange tolerates a fully malicious network at $O(\lambda+\ell)$ communication. The prototype's secure-channel unit implements this hardened tier (Section~\ref{subsec:realizing-security-boundary}). Every headline number uses the default deployment; the anonymous supplementary material details KORD-mac.

\section{Hardware Architecture and Implementation}
\label{sec:hardware-architecture}

All cycle-level results reported here come from cycle-accurate simulation of the same RTL synthesized in Section~VII-E.

\subsection{Realizing the Security Boundary}
\label{subsec:realizing-security-boundary}

The boundary of Section~II-B is enforced by three mechanisms: a single message interface, fail-closed authentication, and special-purpose datapaths that execute no loaded code. The trusted perimeter contains the hardware inside the message interface: the SCU, reconstruction zone, fabric, and assembler. Nothing outside this perimeter is trusted with the counterparty's material. The system is therefore \emph{dealerless under a minimal hardware root}, not trust-free. The root is small, special-purpose, and auditable, but it remains a root.

The SCU is the sole message interface. Every key crosses the boundary in the fixed ciphertext-frame layout of Section~\ref{subsec:one-round-generation}; under KORD-mac the FSS mode and widths ride as associated data, authenticated but not hidden, adding no per-message bytes. The prototype's SCU implements the hardened KORD-mac channel (per-message AES-GCM) and is fail-closed: authentication failure yields no decryption, no fabric dispatch, and no observable engine state (the 20/20 negative-trial evidence is in Section~VI). The default channel replaces the GHASH/verify stage with the session-keystream XOR and is a strict subset of the same hardware, so the SCU figures of Section~VII-E upper-bound the default channel's cost; the engine and fabric behind the SCU are channel-agnostic. The remaining secrets never cross the boundary. The per-key roots are derived on-chip from $K_{\mathrm{ch}}$ through $k_r$, and the pairing step establishes $K_{\mathrm{otp}}$ and the KORD-mac channel key $k_p$. Replay protection requires the sender to maintain a monotonic counter.

Two consequences follow. First, the GGM fabric is realizable only inside the trust boundary: its interleaved expansion state is, level by level, the key material itself, so the same cross-key pipeline built outside the boundary would be the dealer with extra steps. Second, KORD is \emph{not a generic AES accelerator}. Hosts already have AES throughput; the boundary instead protects the joint view of both parties' seed chains to which that throughput is applied.

\subsection{Datapath and Dataflow}
\label{subsec:datapath-dataflow}

\begin{figure*}[t]
  \centering
  \FigureImage{2.85in}{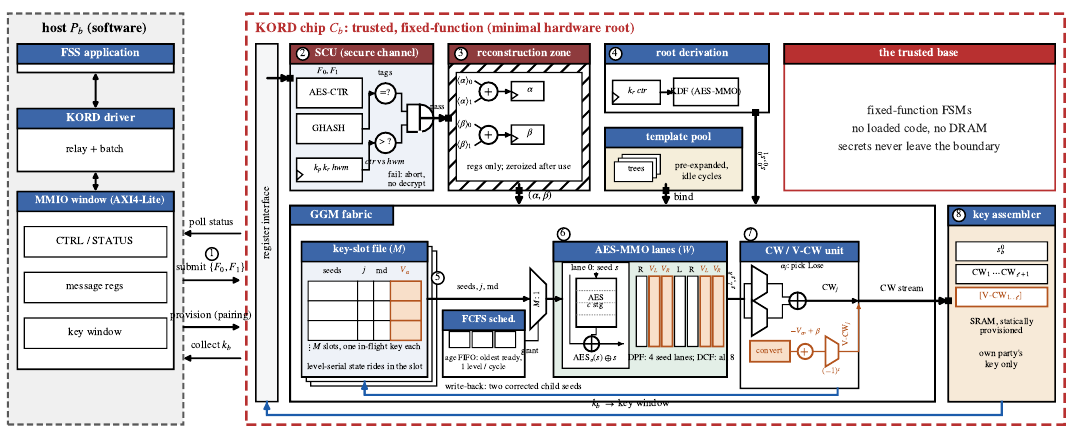}
  \caption{The complete KORD hardware organization. The host software stack is on the left, with four memory-mapped operations as its entire view of the chip. The red dashed trust boundary encloses every hardware element on the key path. Wedged boxes are clocked registers, trapezoids are muxes, $\oplus$ denotes XOR, and the D-shaped gate is the fail-closed dispatch AND. This gate remains closed unless both tags verify and the counter advances. Circled numbers 1--8 trace one key's dataflow. Orange elements (value lanes, the $V_\alpha$ slot field, and the value-correction path) act only for DCF keys. Plaintext $(\alpha,\beta)$ exists only in the hatched reconstruction zone. The gray dashed panel is the semi-honest software side.}
  \label{fig:kord-hardware}
\end{figure*}

Figure~\ref{fig:kord-hardware} shows the host software, the trust boundary, and every hardware module traversed by a key. The host side contains an unchanged FSS application and a driver that accesses the chip through four memory-mapped operations. Inside the boundary, the key passes through the register interface and the SCU. The SCU contains one AES-GCM core with its GHASH, an FSM that authenticates the two messages serially, and the replay check. The remaining modules are the register-level reconstruction zone; the root-derivation unit, which converts $(k_r,\mathit{ctr})$ into the two roots in lockstep with the peer chip; the template pool; the GGM fabric, comprising a key-slot file of $M$ in-flight keys, an FCFS scheduler, $W$ AES-MMO lanes, and the correction-word unit; and the key assembler. Plaintext $(\alpha,\beta)$ exists only in the hatched zone, and every orange element acts only for DCF keys.

\emph{DPF dataflow.} \textcircled{1} The driver writes the key's type parameters, counter, and the two encrypted frames, and starts the operation. \textcircled{2} The SCU authenticates $F_0$ then $F_1$, each tag checked over the associated data (mode, $b_{\mathrm{in}}$, $b_{\mathrm{out}}$, $\mathit{ctr}$) and the counter against the replay high-water mark. Any failure ends the flow with no decryption and no dispatch. \textcircled{3} Only on double authentication does the reconstruction zone latch $\alpha\gets\langle\alpha\rangle_0+\langle\alpha\rangle_1$, $\beta\gets\langle\beta\rangle_0+\langle\beta\rangle_1$ (registers, not SRAM, zeroized once consumed). \textcircled{4} The root-derivation unit supplies $(s_0^0,s_1^0)$ from $(k_r,\mathit{ctr})$. The key enters a free slot at level zero with its mode bit clear. \textcircled{5} Each cycle, the scheduler grants the oldest slot whose previous level has returned and issues the next complete level. A DPF level requires $c=4$ AES-MMO calls: the $L/R$ seed lanes of both chains. \textcircled{6} The four lanes expand with equal latency and return in the same cycle. \textcircled{7} The correction-word unit reads the path bit $\alpha_j$, forms $\mathit{CW}_j$ by Eq.~(3), and writes exactly two corrected child seeds back to the slot; after the last level it forms the final correction by Eq.~(4). \textcircled{8} The assembler serializes $(s_b^0,\mathit{CW}_1,\ldots,\mathit{CW}_{\ell+1})$ in statically provisioned SRAM with no heap or residue. The key leaves through the SCU egress into the key window, where the host collects its own party's share and nothing else.

\emph{DCF dataflow.} A DCF key walks the same eight stations with its mode bit set. This bit changes exactly three components, all orange in Figure~\ref{fig:kord-hardware}. At admission \textcircled{4}, the slot arms its running value accumulator $V_\alpha$. At issue \textcircled{5}, the level uses $c=8$ lanes instead of four. The $V_L/V_R$ value lanes of both chains fire alongside $L/R$, carrying the value-correction chain of Algorithm~\ref{alg:kord-dcf-generation}. At extraction \textcircled{7}, the unit runs its value path after the seed path, forming $\mathit{CW}_j^V$ with the sign fixed by $t_1^{j-1}$ and folding the level into $V_\alpha$, which resides in the slot. The final correction is the converted level-$\ell$ seeds minus $V_\alpha$, and the assembler appends the value corrections to the key layout. No other chip component distinguishes DPF from DCF.

The template pool holds derived-mask trees (trees whose hidden points the chips draw in lockstep from the seed stream, not from any party's input), pre-expanded during idle cycles. Their contents never leave the boundary, and their only public attributes are the type parameters bound as associated data, so a message cannot be replayed against a different template. The direction of the datapath never reverses: no unit downstream of the SCU can emit data except through the SCU egress, and the reconstruction zone is written by the SCU and read by the fabric only.

\subsection{Host Integration: Software--Hardware Co-Design}
\label{subsec:host-integration}

The chip exposes four memory-mapped operations to its host (left half of Figure~\ref{fig:kord-hardware}): provision at pairing time, submit one key's two messages, poll status, and collect the emitted key stream. These operations and their encrypted messages are the entire host-visible surface of the trusted base. The division of labor follows the trust split. Host software performs all flexible work that touches no secret beyond its own party's: assembling and relaying messages, batching submissions to keep $M$ keys in flight, retrying after transport failure, and passing collected keys to the unchanged FSS stack. The chip performs every step that touches a secret beyond the party's own: authentication, reconstruction, expansion, and emission. A driver that corrupts, reorders, or replays ciphertext frames is indistinguishable to the chips from the malicious network and is rejected fail-closed. A driver that alters its party's plaintext shares before encryption deviates from the protocol, which the semi-honest host assumption excludes. No cryptography in the chip can detect this deviation.

The prototype implements this interface as an AXI4-Lite register window on the ZCU102, with Linux on the processing system as the host. Bit-exact comparison against the engine regression's golden vectors verifies the interface, including the fail-closed negative trials.

\subsection{Two Phases: Batch Generation, Constant-Time Bind}
\label{subsec:two-phases}

The datapath uses a split execution model. During the \emph{offline} phase, the fabric batch-expands GGM trees at full utilization and off the critical path. The bind path is narrow by construction because a BGI key's correction words depend on every bit of $\alpha$. Templates therefore serve only derived-mask key classes, whose hidden points are drawn in lockstep from the seed stream before any request. A transmitted-share request takes the direct-generation path. \emph{Bind} updates only payload-dependent corrections: the single final correction word for DPF, independent of $\ell$, and also the per-level value corrections for DCF, linear in $\ell$. The prototype does not implement binding. We therefore withdraw the previously modeled bind-versus-expand ratios; the verified path is direct generation. The prototype verifies the direct-generation path end-to-end. It co-simulates 60 keys through the SCU, reconstruction, fabric, and assembler. It records 4,540 checks and 60/60 authenticated messages with zero errors. The co-simulation implements both chips' identical local generation in one engine. A two-chip deployment runs the same computation independently at each site.

\subsection{The Wall, Measured on the RTL}
\label{subsec:rtl-wall}

In hardware, per-key utilization is bounded by the RTL latency. One expansion level takes $L_{\mathrm{eff}}=12$ cycles: 11 cycles of AES-lane pipeline latency plus one cycle of scheduling turnaround. A single key therefore occupies the array at $1/L_{\mathrm{eff}}=8.31\%$ (measured in cycle-accurate RTL simulation, not modeled), the remaining issue slots being dependency bubbles (Figure~\ref{fig:ggm-fabric}). This floor applies to the simulated $c=W$ configurations: DPF $W=4$ and unified $W=8$. With $c<W$, the same law gives the DPF-only $W=8$ engine a 4.17\% floor and a knee at $M^*=WL_{\mathrm{eff}}/c=24$. All utilization figures below therefore refer to the simulated $c/W=1$ pair. Every reported utilization comes from an engine whose outputs match a software oracle bit-for-bit: 108,276 checked values for the DPF engine and 226,920 for the unified DPF/DCF engine, with zero errors. The AES core also passes the FIPS-197 known-answer tests.

\subsection{Cross-Key Interleaving: $W$ Lanes, FCFS, $M$ Slots}
\label{subsec:cross-key-interleaving}

\begin{figure*}[t]
  \begin{minipage}[t]{0.485\textwidth}
    \centering
    \FigureImage{3.65in}{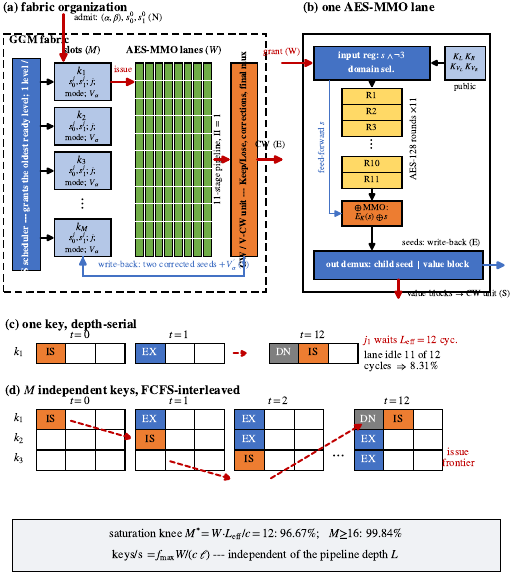}
    \captionof{figure}{The GGM fabric. (a) Organization: the FCFS scheduler grants one ready level per cycle from the $M$ key slots into the $W$-lane AES-MMO array; the correction-word unit consumes the returning blocks and writes two corrected seeds (plus $V_\alpha$) back. (b) One AES-MMO lane: domain-key select, the 11-round AES pipeline (initiation interval 1), the MMO feed-forward XOR, and the output demux. (c),(d) The schedule, cycle by cycle (rows: keys; columns: GGM levels; IS = issuing, EX = in the lanes, DN = level done). One key is depth-serial: $j_1$ waits $L_{\mathrm{eff}}=12$ cycles after $j_0$, idling the lane 11 of every 12 cycles (8.31\%); with $M$ independent keys the issue frontier (red dashed) launches a different ready key every cycle, so at the knee $M^*=12$ the keys tile one $L_{\mathrm{eff}}$ span exactly.}
    \label{fig:ggm-fabric}
  \end{minipage}\hfill
  \begin{minipage}[t]{0.485\textwidth}
    \centering
    \FigureImage{2.35in}{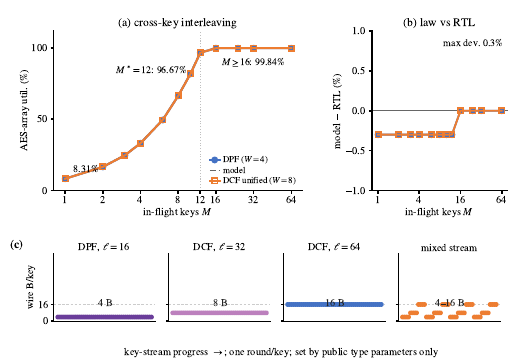}
    \captionof{figure}{(a) Utilization of the shared AES array vs in-flight keys $M$, cycle-accurate RTL simulation (DPF $W=4$ and unified DCF $W=8$, both $c/W=1$, so the saturation law predicts one curve and the RTL confirms it): single chain 8.31\%; knee $M^*=12$ reaches 96.67\%; $M\ge16$ sustains 99.84\%. (b) The closed-form law deviates from the RTL by at most 0.3\% at every $M$. (c) The exchange segment's observable surface: wire bytes per key are the fixed $2\lceil\ell/8\rceil$-byte frame (KORD-mac: +36~B) and one round, set by the public type parameters alone (design values; message-level observables only, physical side channels not empirically evaluated).}
    \label{fig:utilization-sweep}
  \end{minipage}
\end{figure*}

The fabric fills each dependency bubble with a ready level from another key. Figure~\ref{fig:ggm-fabric} shows its organization, one lane's microarchitecture, and the cycle-by-cycle schedule. The $W$ AES lanes accept one block per cycle (initiation interval 1). The $M$ key slots hold each key's current seed pair, level counter, and correction-word accumulator. Each cycle, a first-come-first-served scheduler issues the oldest slot whose next level is ready. When a key stalls on its own dependency, another ready key fills its bubble (Figure~\ref{fig:ggm-fabric}d). Utilization therefore depends on the number of independent chains in flight, not on the latency of one chain.

Saturation has a closed form. The array is full when $M^*=W\cdot L_{\mathrm{eff}}/c$ keys are in flight, where $c$ is the number of AES calls per level. The swept configurations have $M^*=12$. Past this knee, the saturated rate is $\text{keys/s}=f_{\max}W/(c\ell)$, with no $L$ term. Pipeline depth moves the knee rather than the throughput frontier. AES latency thus sets the number of in-flight keys required for saturation, not the output rate. Figure~\ref{fig:utilization-sweep} sweeps $M$ in cycle-accurate RTL simulation: at the knee $M=M^*$ the array reaches 96.67\% (pipeline utilization, not to be conflated with the numerically adjacent key-generation time share $\ge96.9\%$ of Section~VII-C), and four slots of headroom close the remainder: $M\ge16$ sustains 99.84\%. At $M=M^*$, the scheduler has no redundant slots. A phase collision between two ready keys therefore wastes an issue slot that FCFS cannot recover. Four additional slots, requiring a few hundred bytes of state, eliminate this loss. The analytical model tracks the measured curve within 0.3\% at every swept point, so the closed form can provision the fabric.

\emph{Storage:} The implemented in-flight state is 52.6~B per key, so the saturated working set at $M=16$ is 0.82~KiB. The analytical model allocates 64~B/key; the implementation uses less because keys stream out without a landing buffer.

\subsection{Calibers, the Frequency Limit, and One Datapath}
\label{subsec:frequency-limit}

We report two absolute quantities. First, cross-key interleaving raises utilization of the same hardware from 8.31\% to 99.84\%. Second, at equal throughput, a purely data-parallel design with $L_{\mathrm{eff}}$ independent single-chain engines would require 12 copies of the AES array to match one interleaved fabric and its 0.82~KiB of slot state. The fabric therefore delivers $\approx12\times$ the throughput per unit of AES area; the baseline area is an arithmetic extrapolation, not a synthesized design. We do not report speed-up as a headline result. The measured $12.01\times$ saturation speed-up is bounded by $L_{\mathrm{eff}}$ by construction and only validates the model against the RTL.

The AES lanes do not limit frequency. Synthesized alone against a tight 2.5~ns target, both lanes close timing with positive slack. Their WNS-extrapolated frequencies are 409.0~MHz for the DPF lane and 420.2~MHz for the unified-FSS lane. In contrast, the full engines' 204.0 and 154.6~MHz values are WNS-extrapolated $f_{\max}$ figures under $f_{\max}=1000/(\text{period}-\text{WNS})$, not timing-closure frequencies. The frequency gap comes from the correction-word extraction logic. This combinational reduction folds a level's lane outputs into a correction word and remains unpipelined in the current design. Pipelining it would raise $f_{\max}$ and, by the saturation law, require more slots without reducing rate.

\emph{One datapath for DPF and DCF.} DCF generation adds a value-correction-word chain, but unifying DPF and DCF under one scheduler does not require a second datapath. Post-route, the unified engine uses $+0.96\%$ LUTs over a DCF-only datapath (20,654 vs 20,457) at $-0.25\%$ $f_{\max}$ (154.6 vs 154.9~MHz). Replacing two separate engines, one DPF and one DCF, saves about half the resources: 50.8\% of LUTs and 50.3\% of BRAMs (41,997 LUT/1,173 RAMB18 for the pair vs 20,654/583 unified). All figures are ZCU102 place-and-route estimates (synthesis-estimated, not board measurements). The scheduler already treats levels as independent issue units. A DCF level uses $c=8$ AES calls instead of 4, and the saturation law accounts for this difference through $c$.

\section{Security Analysis}
\label{sec:security}

Under the threat model of Section~\ref{subsec:threat-model}, KORD is dealerless under a minimal hardware root: no principal ever plays dealer. The anonymous supplementary material gives the full proofs.

\textbf{Qualitative analysis.}
For each party $P\in\{A,B\}$, the joint view of an adversary controlling $P$ (semi-honest), the host at $P$'s site, and the network (semi-honest by default; malicious under KORD-mac) is computationally simulatable from $P$'s own shares $(\alpha_P,\beta_P)$, the public template parameters, the message count, and message timing. The view is thus computationally independent of the true $(\alpha,\beta)$ and the honest party's shares; even the delivered key share is simulated from single-share FSS pseudorandomness (Lemma B.3). The proof composes three simulators: pairing (authenticated key establishment and attestation, imported as primitives); the exchange (fixed-length ciphertexts of uniformly distributed share strings, Figure~\ref{fig:utilization-sweep}c---one keystream block per (counter, direction) message makes the default channel information-theoretic given counter freshness, while KORD-mac reduces to authenticated encryption plus counter freshness); and the chip (an ideal functionality for reconstruction, template binding, and root-seed derivation---the stated hardware-root assumption---whose emitted shares are individually pseudorandom). The claim is bounded in two ways: it is conditional on the trusted-chip abstraction, so minimality denotes the bounded decomposition, not a proved lower bound; and across requests it reduces to multi-instance PRF security of the derivation from $K_{\mathrm{ch}}$, since single-key FSS security does not cover the correlated stream.

\textbf{Against a dealer in a box.}
No principal sees plaintext: a dealer---software with an operator and an update channel---handles every $(\alpha,\beta)$ in the clear; KORD's plaintext exists only in a register-level zone, zeroized after use. The trusted object is the special-purpose hardware and authenticated channel of Section~\ref{subsec:realizing-security-boundary}---34,199 LUT and 663 RAMB18 (synthesis-estimated, Section~\ref{sec:hardware-eval})---rather than a general-purpose stack. Little of its work is input-dependent: batch GGM expansion touches only request-independent templates (Section~\ref{subsec:two-phases}); only reconstruction, binding, and emission touch secrets; the verified direct-generation path stays inside the same boundary. Minimal denotes the three-segment decomposition, not a proved minimum: KORD relocates rather than eliminates trust and retains the stated cross-party cost.

\textbf{The network and the boundary.}
The default tier assumes integrity: relays may read everything but deliver faithfully, confidentiality is the keystream's, and a keystream mask is malleable, so a tampering relay falls outside the model rather than being defeated by the channel. KORD-mac enforces it: the chips accept no unauthenticated message from any source, and the SCU is fail-closed---a failed message yields no decryption, no engine dispatch, and no observable engine state. All 20 negative trials (tampered tag, tampered ciphertext, wrong channel key, replay) fail with zero engine-input leakage (20/20), and the replay guard advances its high-water mark only after a message proves authentic and fresh. Neither tier enforces availability: a network that withholds a request from one chip desynchronizes the counters, and the affected key pairs become unusable rather than leaked; this is out of scope, with counter re-alignment a host-visible resynchronization step. Direction: distinct per-direction subkeys and keystreams (the direction is in the KDF domain), so a reflected or cross-substituted frame decrypts to garbage or fails authentication. Persistence: counters and pairing secrets must reside in rollback-protected storage (the FPGA prototype implements only in-session monotonicity). A rollback is a security failure, not only a correctness one---a reused keystream pad leaks the XOR of two messages, and re-derived roots break the one-time-key guarantee---so KORD requires transactional persistence with counter commit before emission, idempotent re-emission, and one recovery rule: an advanced counter is never rolled back; the lagging chip advances past skipped roots without emitting keys.

\textbf{Side channels.}
The register-level, zeroize-after-use design of the reconstruction zone is a by-construction mitigation only; we have performed no empirical side-channel evaluation and make no claim of leakage freedom.

\section{Evaluation}
\label{sec:evaluation}

We evaluate KORD against deployed key sources, from per-operation cost to full-system key supply.

\subsection{Experimental Setup and Methodology}

\textbf{Synthesis platform.}
All hardware figures target a single Xilinx Zynq UltraScale+ MPSoC ZCU102 board, device XCZU9EG-2FFVB1156, using Vivado 2022.2. Each engine, the secure-channel unit, and the full online path are synthesized and placed-and-routed out-of-context against a common timing target; $f_{\max}$ is extrapolated from the post-route worst negative slack as $f_{\max}=1000/(\mathrm{period}-\mathrm{WNS})$, and power is a vectorless estimate at default switching activity. These area, RAMB18, and power figures are synthesis estimates; nothing has been taped out.

\textbf{Board validation.}
The same online-path design---secure-channel unit, reconstruction zone, and unified engine behind an AXI4-Lite window---is programmed onto the board and exercised on silicon. We drive the AXI window through the processing system's JTAG debug-access port rather than a host operating system: \texttt{psu\_init} configures the PS clocking and the \texttt{M\_AXI\_HPM0\_FPD} master, the bitstream is loaded over JTAG, and the register reads and writes that submit a key and collect its output are issued over the debug link. Because this register path never touches off-chip DRAM, board validation needs no operating system, storage, or working memory. Board $f_{\max}$ is measured by retuning the PL clock in place across 100--250~MHz---each setting confirmed by reading back the clock divider---and re-running the golden vectors; saturated throughput and lane utilization are measured with an on-chip request generator that keeps the engine's in-flight slots full.

\textbf{RTL and software references.}
Key correctness is established in Vivado xsim against a C++ golden model linked to our \texttt{fss-ref} reference implementation, and the same golden vectors are replayed on the board. All baseline numbers re-run published artifacts as genuine two-party deployments on the same host. The dealerless generator of Xing et al.~\cite{ref54} is recompiled and run as two simultaneous processes on localhost, reading the framework's own byte and round counters (median of three); MP-SPDZ~\cite{ref27} is its semi2k offline phase at ring width $R=64$, the smallest its reference binaries support; and the trusted-dealer LLAMA stack~\cite{ref22} anchors the end-to-end comparison.

\textbf{Workloads.}
We use nine FSS building blocks at $\ell\in\{8,16,18\}$ (localhost, median of three), private ResNet-18~\cite{ref23} on ImageNet ($\ell=32$, scale 10) over a 1~Gbps LAN, the baseline paper's own 40~Mbps, 100~ms-RTT WAN setting, and distributed ORAM at $d=16$.

\textbf{Provenance and calibers.}
Every number carries one of three labels: measured (re-run artifacts), formula-extrapolated (a baseline's own complexity formula beyond its code's range), or synthesis-estimated (KORD values from the platform above). KORD's message sizes come from frozen byte-level accounting, and we never place KORD's model latency in a speed-up ratio against a measured value; before timing, we removed one unconditional debug print from the baseline's timing region, affecting neither logic nor byte counts. The dealerless baseline is reported under two calibers: caliber A, its own communication formula ($9\ell+5$ rounds and $18\lambda\ell+5\lambda+13\ell+2$ bits per DPF key, $16\ell+7$ rounds per DCF key, $\lambda=128$), a lower bound favorable to the baseline; and caliber B, our reproduced measurement (79.6~KB per party per DPF key and 276.8~KB per DCF key at $\ell=16$). The artifact crashes above $\ell=18$, so all wider figures are labeled extrapolations (A by formula, B by a linear fit whose ratio to A is stable on the measured grid). Party-side EvalAll traffic is not modeled, since KORD changes generation, not evaluation.

\textbf{Plan.}
Sections~\ref{sec:building-blocks}--\ref{sec:hardware-eval} answer, in turn: per-operation cost across the nine building blocks and its scaling with $\ell$; system-level impact on private ResNet-18; behavior beyond the LAN, on the WAN and for distributed ORAM; and the silicon cost and key rate of the synthesized engines.

\subsection{Building Blocks and Input-Width Scaling}
\label{sec:building-blocks}

Table~\ref{tab:blocks} reports per-operation generation cost for the nine FSS building blocks used by a private-inference and lookup stack. One operation may consume several primitive keys, and each column totals the block's complete key complement. We use $\ell\in\{8,16,18\}$, the widest range supported by the baseline code. KORD reduces per-key communication by 7,633$\times$ to 70,274$\times$ (caliber B) and reduces interaction from 154--38,659 one-way messages to a single round. KORD's own key message is 2--512 bytes. KORD's generation latency across all these cells is 0.45--5.49~$\mu$s. This synthesis-estimated model value shows only that generation keeps pace; it is not a speed-up. The baseline's measured localhost generation time is 37.8~ms--7.14~s.

A second dealerless reference, MP-SPDZ (semi-honest 2PC, dishonest majority), is compared on its offline (key-generation) axis. Its preprocessing communication exceeds KORD's by 434$\times$ to 263,940$\times$. A batched comparison still amortizes to 1,535~B/op, 384$\times$ KORD's message at $\ell=8$. Every MP-SPDZ number uses ring width $R=64$, the smallest supported by its reference binaries.

\begin{table*}[t]
\caption{Per-operation generation cost across nine FSS building blocks at $\ell\in\{8,16,18\}$; each column totals the block's full key complement, per party per operation. Xing~\cite{ref54} and MP-SPDZ~\cite{ref27} columns are measured (localhost, median of 3); KORD columns are a synthesis-estimated message/cycle model, and the ratio is caliber B (Xing/KORD).}
\label{tab:blocks}
\centering
\footnotesize
\setlength{\tabcolsep}{4pt}
\begin{tabular}{lrrrrrrrr}
\toprule
& & \multicolumn{3}{c}{Xing dealerless (measured)} & \multicolumn{2}{c}{MP-SPDZ off. ($R=64$)} & \multicolumn{2}{c}{KORD (model)}\\
\cmidrule(lr){3-5}\cmidrule(lr){6-7}\cmidrule(lr){8-9}
BB & $\ell$ & Comm. (B) & Msgs & Gen (ms) & Comm. (B) & Rnds & Comm. (B) & Ratio\\
\midrule
EQ     & 8  & 46,283    & 154    & 37.8   & 8,425 & 18 & 2   & 23,142$\times$\\
EQ     & 16 & 79,628    & 298    & 69.6   & 8,425 & 18 & 4   & 19,907$\times$\\
CMP    & 8  & 281,098   & 551    & 123.6  & 7,393 & 18 & 4   & 70,274$\times$\\
CMP    & 16 & 545,272   & 1,079  & 274.0  & 7,393 & 18 & 8   & 68,159$\times$\\
CMP    & 18 & 611,496   & 1,211  & 385.0  & 7,393 & 18 & 12  & 50,958$\times$\\
PriLUT & 8  & 9,596,288 & 38,659 & 7,142.8& --    & -- & 512 & 18,743$\times$\\
Approx & 8  & 1,179,844 & 4,151  & 804.7  & --    & -- & 100 & 11,798$\times$\\
Approx & 18 & 1,511,278$^{\dagger}$ & 4,811 & 960.7 & -- & -- & 198 & 7,633$\times$\\
\bottomrule
\end{tabular}

\vspace{2pt}\parbox{0.98\textwidth}{\scriptsize Aggregate wire traffic sums both parties' volumes; for KORD, it is twice the printed per-party value, and the hardened KORD-mac tier adds a fixed 36~B per message. Xing's Msgs column counts one-way messages; MP-SPDZ's Rnds is its reported offline round count. Rows are a representative subset of the $9\times3$ grid covering both ratio endpoints (7,633$\times$, 70,274$\times$); the omitted blocks (Mod, TR, Ctn, DigDec, PubLUT) fall between them, with Mod matching CMP on every cost column. MP-SPDZ offline cost is identical at every $\ell$ of a block because it is set by ring width (F5). ``--'' marks blocks with no faithful semi2k microbenchmark (F2). One-time OT setup is amortized across a block's keys (F1). $\dagger$ denotes an artifact measurement. The baseline's paper prints 1.151~MB, breaking its own monotone trend, attributed to a digit transposition (E1).}
\end{table*}

One cell is inconsistent with the baseline's trend. Its paper prints 1.151~MB for the $\ell=18$ approximation gadget, breaking its own monotone trend (1.180~MB at $\ell=8$, 1.445~MB at $\ell=16$), while the artifact measures 1,511,278~B. We attribute the printed value to a digit transposition and report our measurement.

\begin{figure}[t]
\centering
\FigureImage{3.00in}{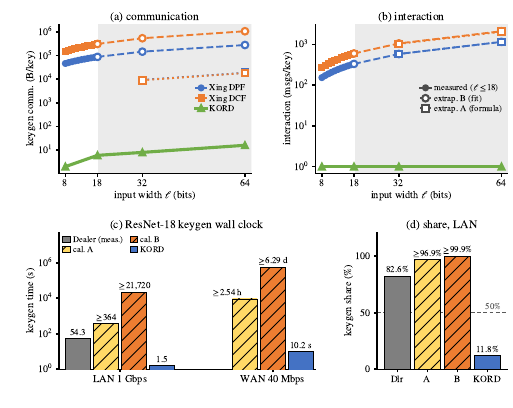}
\caption{(a),(b) Input-width scaling of per-key generation: communication and interaction for DPF and DCF vs. $\ell$. Baseline measured up to $\ell=18$ (filled); the shaded region is caliber B (linear fit, open circles) and caliber A (complexity formula, open squares); KORD is flat at $2\lceil\ell/8\rceil$~B and one round. (c),(d) The ResNet-18 key-stream experiment: key-generation wall-clock time on the LAN (1~Gbps) and the baseline's own 40~Mbps WAN setting (log scale), and key generation's share of end-to-end time on the LAN. Hatched bars are extrapolated lower bounds with the baseline's generation compute and tree memory excluded (favorable to the baseline). The dealer bar is measured, whereas KORD bars are synthesis-estimated model totals. No dealer deployment is defined for the WAN scenario. KORD is the only deployment below the 50\% line: the online phase re-dominates.}
\label{fig:scaling}
\end{figure}

\textbf{Input-width scaling.}
The input-width sweep isolates asymptotic scaling. Across the measured grid ($\ell\in\{8,\ldots,18\}$) the baseline's key-generation communication is strictly linear in $\ell$: 4,168.2~B/bit for DPF and 16,500.7~B/bit for DCF, with maximum residuals of 0.00\% and 0.01\%; the one-way message count is twice the formula rounds for DPF and $33\ell+9$ for DCF. When both calibers are extrapolated to production widths (Figure~\ref{fig:scaling}, shaded region), per-key DPF communication exceeds KORD's by 1,169$\times$ (caliber A) to 18,290$\times$ (caliber B) at $\ell=32$. The DCF gap at this width is 1,132$\times$ to 67,611$\times$; at $\ell=64$, it reaches 1,146$\times$ (A) to 66,807$\times$ (B). KORD's message remains nearly flat (4 $\rightarrow$ 16~B), and its round count remains one.

We report the baseline's generation latency only as a lower bound ($\geq147.7$--701.7~ms across the extrapolated widths). Its Gen protocol expands the full GGM tree per key, requiring $\Theta(2^\ell)$ PRG evaluations under Algorithm~2. Any latency figure omits this cost and is therefore a lower bound by construction.

\textbf{Direct comparison on the baseline's own tables.}
Table~\ref{tab:baseline} reproduces the baseline's key-generation measurements (its Tables II and III) and adds KORD under the identical network settings, on the generation axis alone. Two effects separate. Generation communication collapses from the baseline's 0.046--9.6~MB per operation to KORD's 2--198~B, a 7,633$\times$--72,500$\times$ reduction, because KORD publishes one correction-word share per level instead of an interactive OT transcript. Generation latency collapses because KORD is a single round: the baseline's $\Theta(\ell)$-round protocol costs 0.05--2.16~s on the LAN and 1.3--326~s on the WAN and grows with both $\ell$ and protocol complexity, whereas KORD's one RTT pins its latency near 0.05~ms and 20~ms regardless of protocol, yielding 1,000$\times$--43,140$\times$ (LAN) and 64$\times$--16,312$\times$ (WAN) reductions. The WAN gap widens with protocol complexity precisely because round count, not bytes, drives the baseline there. Online evaluation is excluded by construction: KORD emits standard keys, so the parties' Eval phase is bit-identical to the baseline's and to a dealer's.

\begin{table*}[t]
\caption{Direct reproduction of the baseline's own key-generation measurements~\cite{ref54} (its Tables II and III), restricted to the generation axis that KORD changes. The two upper rows are the DPF/DCF primitives; the lower block is its FSS building blocks. Xing columns are its paper's reported values, measured on AWS C5.9xlarge at 3.6~GHz over the same two links it uses: a 10~Gbps/0.05~ms LAN and a 40~Mbps/20~ms WAN. Online evaluation is not compared: KORD emits standard keys, so its Eval is bit-identical to the baseline's. KORD figures are for KORD's own environment (Section~VII-A): a single-round protocol whose generation latency is one RTT plus a 0.34--5.49~$\mu$s fabric expansion (synthesis-estimated, board-confirmed rate law, Section~VII-E) plus a 2--512~B transfer. Because one RTT dominates, KORD's generation latency is pinned near 0.05~ms (LAN) and 20~ms (WAN) independent of protocol and $\ell$; the reduction columns are against these. $\dagger$ Artifact measurement; the baseline's paper prints 1.151~MB, a digit transposition breaking its own trend (E1).}
\label{tab:baseline}
\centering
\scriptsize
\setlength{\tabcolsep}{3.2pt}
\begin{tabular}{lrrrrrrrr}
\toprule
& & \multicolumn{2}{c}{Gen, Xing (s)} & \multicolumn{2}{c}{Comm., generation} & \multicolumn{3}{c}{KORD reduction}\\
\cmidrule(lr){3-4}\cmidrule(lr){5-6}\cmidrule(lr){7-9}
Protocol & $\ell$ & LAN & WAN & Xing & KORD & Gen LAN & Gen WAN & Comm.\\
\midrule
$\Pi_{\mathrm{DPF}}$ & 8  & 0.050 & 1.288 & 0.046 MB & 2 B & 1,000$\times$ & 64$\times$ & 23,142$\times$\\
$\Pi_{\mathrm{DPF}}$ & 18 & 0.124 & 2.780 & 0.088 MB & 6 B & 2,480$\times$ & 139$\times$ & 14,661$\times$\\
$\Pi_{\mathrm{DCF}}$ & 8  & 0.085 & 2.417 & 0.145 MB & 2 B & 1,700$\times$ & 121$\times$ & 72,500$\times$\\
$\Pi_{\mathrm{DCF}}$ & 18 & 0.280 & 5.247 & 0.310 MB & 6 B & 5,600$\times$ & 262$\times$ & 51,667$\times$\\
\midrule
$\Pi_{\mathrm{EQ}}$ & 8  & 0.050 & 1.277 & 0.046 MB & 2 B & 1,000$\times$ & 64$\times$ & 23,142$\times$\\
$\Pi_{\mathrm{EQ}}$ & 18 & 0.125 & 2.783 & 0.088 MB & 6 B & 2,500$\times$ & 139$\times$ & 14,661$\times$\\
$\Pi_{\mathrm{CMP}}$ & 8  & 0.162 & 4.819 & 0.281 MB & 4 B & 3,240$\times$ & 241$\times$ & 70,274$\times$\\
$\Pi_{\mathrm{CMP}}$ & 18 & 0.539 & 10.605 & 0.611 MB & 12 B & 10,780$\times$ & 530$\times$ & 50,958$\times$\\
$\Pi_{\mathrm{TR}}$ & 8  & 0.108 & 3.127 & 0.182 MB & 4 B & 2,160$\times$ & 156$\times$ & 45,537$\times$\\
$\Pi_{\mathrm{TR}}$ & 18 & 0.199 & 5.947 & 0.347 MB & 8 B & 3,980$\times$ & 297$\times$ & 43,401$\times$\\
$\Pi_{\mathrm{Ctn}}$ & 8  & 0.608 & 18.891 & 1.099 MB & 16 B & 12,160$\times$ & 945$\times$ & 68,683$\times$\\
$\Pi_{\mathrm{Ctn}}$ & 18 & 2.157 & 42.296 & 2.421 MB & 48 B & 43,140$\times$ & 2,115$\times$ & 50,427$\times$\\
$\Pi_{\mathrm{DigDec}}$ & 8  & 0.130 & 3.821 & 0.203 MB & 6 B & 2,600$\times$ & 191$\times$ & 33,822$\times$\\
$\Pi_{\mathrm{DigDec}}$ & 18 & 0.241 & 7.311 & 0.389 MB & 12 B & 4,820$\times$ & 366$\times$ & 32,404$\times$\\
$\Pi_{\mathrm{PubLUT}}$ & 8  & 0.050 & 1.282 & 0.046 MB & 2 B & 1,000$\times$ & 64$\times$ & 23,142$\times$\\
$\Pi_{\mathrm{PubLUT}}$ & 18 & 0.126 & 2.778 & 0.088 MB & 6 B & 2,520$\times$ & 139$\times$ & 14,661$\times$\\
$\Pi_{\mathrm{Approx}}$ & 8  & 1.071 & 326.247 & 9.596 MB & 100 B & 21,420$\times$ & 16,312$\times$ & 11,798$\times$\\
$\Pi_{\mathrm{Approx}}$ & 18 & 1.391 & 41.854 & 1.44$^{\dagger}$ MB & 198 B & 27,820$\times$ & 2,093$\times$ & 7,633$\times$\\
\bottomrule
\end{tabular}
\end{table*}

\subsection{End-to-End: Private ResNet-18}
\label{sec:resnet}

We test system-level impact using one private ResNet-18 inference on ImageNet with $\ell=32$ and scale 10. All three deployments compute the same result. KORD's keys drive the online phase to a 1000/1000 bit-exact match against a cleartext reference ($\max|\mathrm{diff}|=0$). The inference consumes 6,726,608 FSS trees. Of these, 5,019,112 form a DCF-equivalent lower bound, using one DCF-Gen@32 per non-linear key.

The baseline's bill is computed in its favor: we exclude the second correction-word tree of truncation and public division, MaxPool multiplications, all Boolean-to-arithmetic and millionaire keys, every convolution and matrix-multiply Beaver triple, and its own key-generation computation and tree-expansion memory. All of these costs are counted as zero. The accounting uses symmetric exclusions: KORD's total also excludes host-side key-delivery I/O. Both columns are therefore compute-and-input-communication models, not delivered end-to-end measurements. Even with these exclusions, key generation on a 1~Gbps unified link is 54.3~s under the trusted-dealer LLAMA stack~\cite{ref22}, or 82.6\% of end-to-end time (measured; Figure~\ref{fig:scaling}c). Under the dealerless baseline, it is $\geq363.6$~s and $\geq96.9\%$ (caliber A). Caliber B, fit-extrapolated to $\ell=32$, raises these values to $\geq21{,}720$~s and $\geq99.9\%$. KORD requires 1.542~s, or 11.9\%. This value is a synthesis-estimated serialized total with generation and transfer summed without overlap credit. It is an upper bound within a model that excludes host-side key-delivery I/O. All three deployments share the same online phase because they emit standard keys. This phase takes 11.43~s over 60.6~MB and 66 rounds (30-repetition median). Key generation dominates end-to-end time under both existing routes. With KORD, the online phase again dominates at 88.1\% (Figure~\ref{fig:scaling}d), restoring the cost profile assumed by online optimizations of the last decade.

\subsection{Beyond the LAN: WAN and Distributed ORAM}
\label{sec:beyond-lan}

On a wide-area link, round count dominates and the gap widens. We use the 40~Mbps application-WAN setting adopted by the baseline's own paper. For a single DCF key at $\ell=32$ over a 100~ms round trip, the baseline-to-KORD latency ratio is 519$\times$ (caliber A) to 520$\times$ (caliber B). This round-count ratio is essentially flat across bandwidth and RTT and rises to 1,031$\times$ at $\ell=64$ (Figure~\ref{fig:beyond}).

Scaled to the ResNet-18 key stream, again granting the baseline its most favorable 519-round batched schedule, dealerless generation takes $\geq2.54$ hours (caliber A) to $\geq6.29$ days (caliber B). KORD takes 10.2~s, a baseline-to-KORD ratio of 898$\times$ to 53,346$\times$ at 100~ms RTT (Figure~\ref{fig:scaling}c). As in the LAN case, these WAN latencies count only communication. The baseline's $\Theta(2^\ell)$ tree-expansion computation and memory are uncharged, so its figures are lower bounds.

\begin{figure*}[t]
\centering
\FigureImage{4.05in}{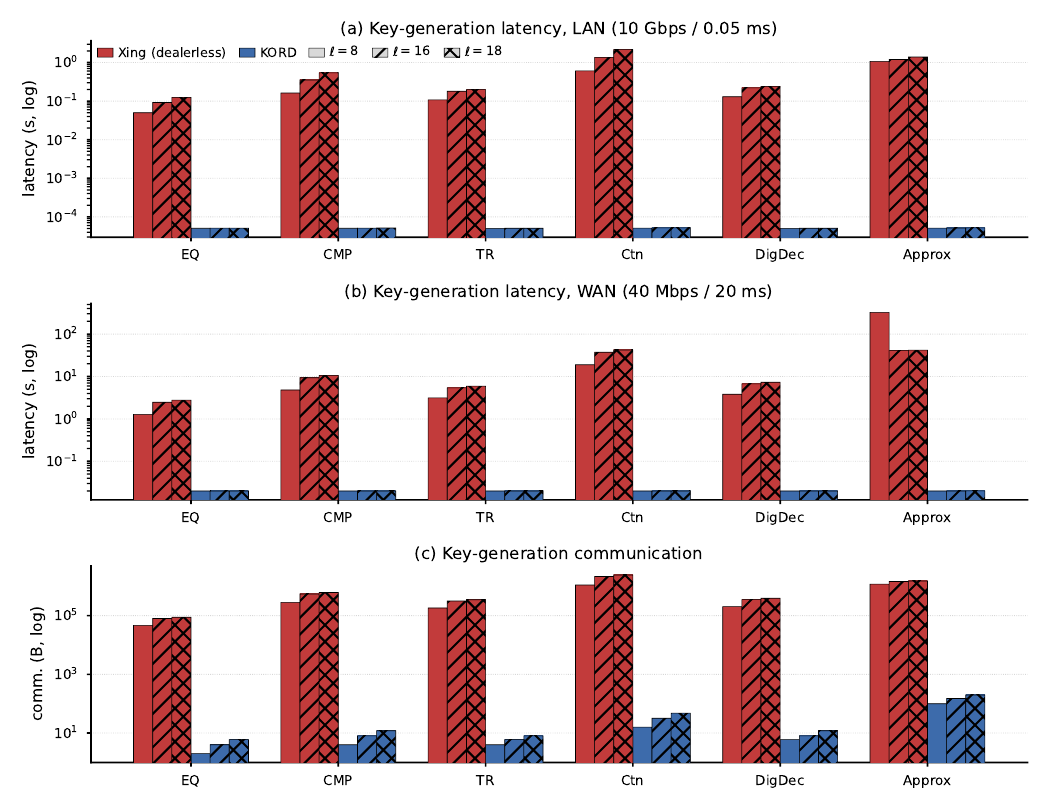}
\caption{Key-generation cost across six FSS building blocks at $\ell\in\{8,16,18\}$, Xing~\cite{ref54} versus KORD, on the generation axis. This mirrors the baseline's own Figure 6---same three panels (LAN time, WAN time, communication)---but on generation rather than online evaluation: the online Eval phase is unchanged by KORD (standard keys), so an online plot would show the two systems coincide, whereas generation is what KORD restructures. Xing bars grow with $\ell$ and protocol complexity ($\Theta(\ell)$ rounds); KORD's single round pins its latency near 0.05~ms (LAN) and 20~ms (WAN) and its message at 2--198~B, independent of both. Log scale.}
\label{fig:blocks}
\end{figure*}
\FloatBarrier

\textbf{Compute versus communication.}
Table~\ref{tab:resnet-breakdown} separates the two overheads the co-design attacks, across both links. Generation compute is network-independent: the trusted dealer spends 18.54~s expanding the 6.73~M GGM trees, whereas KORD's fabric cycle-accounts to 1.14~s---a $\geq16.3\times$ reduction, and a lower bound because the board's measured $f_{\max}$ ($\geq214$~MHz) exceeds the 154.56~MHz used for the estimate (Section~\ref{sec:hardware-eval}). The dealerless baseline charges no compute at all---its $\Theta(2^\ell)$ expansion is left uncharged in its favor---so against it KORD's compute advantage is larger still, and unquantified. Generation communication is where the round-collapse dominates. On the LAN, KORD's single-round 50.4~MB key stream cuts distribution by 88.7$\times$ against the dealer---which must ship 4.47~GB of key material to two remote parties---and by 902$\times$ (A) to 53,896$\times$ (B) against the dealerless baseline. The gap holds on the WAN, at 87.8$\times$ against the dealer and 898$\times$ to 53,346$\times$ against the baseline, because there the baseline's per-key round count multiplies against the 100~ms RTT while KORD stays at one round. The two axes are complementary: the hardware root removes the compute wall ($\geq16\times$) and the single-round protocol removes the communication wall ($10^1$--$10^4\times$), on both the LAN and the WAN.

\begin{table}[t]
\caption{ResNet-18 key generation decomposed into compute and communication overhead, evaluated against the trusted-dealer LLAMA stack~\cite{ref22} and the dealerless baseline~\cite{ref54} (calibers A/B) on both the 1~Gbps LAN and the 40~Mbps/100~ms WAN. Compute is network-independent, so it appears once. The dealerless baseline charges zero compute (its $\Theta(2^\ell)$ tree expansion is uncharged, in its favor), so only its communication is comparable (``--''). LLAMA's compute is measured and its key distribution is bandwidth-scaled to each link. KORD's compute is a cycle-accounted model whose rate law and $f_{\max}$ the board corroborates as conservative (Section~\ref{sec:hardware-eval}), so every KORD-side factor is a lower bound.}
\label{tab:resnet-breakdown}
\centering
\scriptsize
\setlength{\tabcolsep}{3.5pt}
\begin{tabular}{lrrrr}
\toprule
Overhead & LLAMA & Xing A & Xing B & KORD\\
\midrule
Compute (s) & 18.54 & -- & -- & 1.14\\
Comm., LAN (s) & 35.75 & 363.6 & 21,720 & 0.40\\
Comm., WAN (s) & 893.8 & 9,144 & 543,456 & 10.2\\
\bottomrule
\end{tabular}
\end{table}

\begin{figure*}[t]
  \begin{minipage}[t]{0.485\textwidth}
    \centering
    \FigureImage{2.40in}{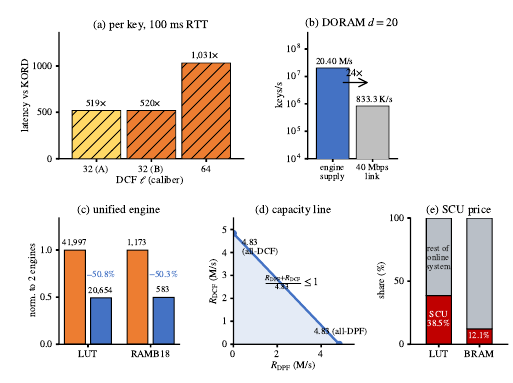}
    \captionof{figure}{Beyond the LAN, and the hardware. (a) Per-key WAN generation-latency ratio (baseline over KORD) at 100~ms RTT, a round-count ratio flat in bandwidth and RTT; caliber in parentheses, hatched = extrapolated. (b) Distributed ORAM at $d=20$: modeled engine supply versus the drain of a 40~Mbps link. (c)--(e) The shared datapath, all synthesis-estimated: unified DPF/DCF engine resources normalized to two separate engines (also within +0.96\% LUT of a DCF-only datapath); the unified engine's mixed-workload capacity line between its two pure endpoints; and the SCU's share of the online system, the hardware price of the trust boundary.}
    \label{fig:beyond}
  \end{minipage}\hfill
  \begin{minipage}[t]{0.485\textwidth}
    \captionof{table}{Synthesized key-generation engines on ZCU102 (XCZU9EG-2); $f_{\max}$/area/power are synthesis-estimated (G2), throughput (DPF-32 keys/s) is board-confirmed (Section~\ref{sec:hardware-eval}). The unified engine issues one level per cycle regardless of mode, so its all-DPF and all-DCF rates are both 4.83~M/s and a mixed stream satisfies $(R_{\rm DPF}+R_{\rm DCF})/4.83\leq1$; the DPF-only engine's 4-lane groups retire two levels per cycle, giving its higher rate. Its keys/J value is for the unified engine and denotes energy per DPF-32-equivalent key. The DPF-only and unified W8 tiers are not interchangeable. $\dagger$ W16 is a capacity boundary where S-boxes partially fold into LUTs, not a linear-scaling point (G5).}
    \label{tab:engines}
    \centering
    \scriptsize
    \setlength{\tabcolsep}{3.1pt}
    \begin{tabular}{lrrrrr}
      \toprule
      Engine & $f_{\max}$ & LUT & RAMB18 & Thpt & keys/J\\
      & (MHz) & & & (M/s) & (M)\\
      \midrule
      DPF W4 & 234.1 & 12,906 & 305 & 7.32 & 2.99\\
      DPF W8 & 204.0 & 21,540 & 601 & 12.75 & 2.75\\
      DPF W16$^{\dagger}$ & 207.8 & 72,092 & 912 & 25.98 & 2.70\\
      Unified W8 & 154.6 & 20,654 & 583 & 4.83 & 2.36\\
      \bottomrule
    \end{tabular}
  \end{minipage}
\end{figure*}

\begin{figure*}[t]
  \centering
  \FigureImage{2.8in}{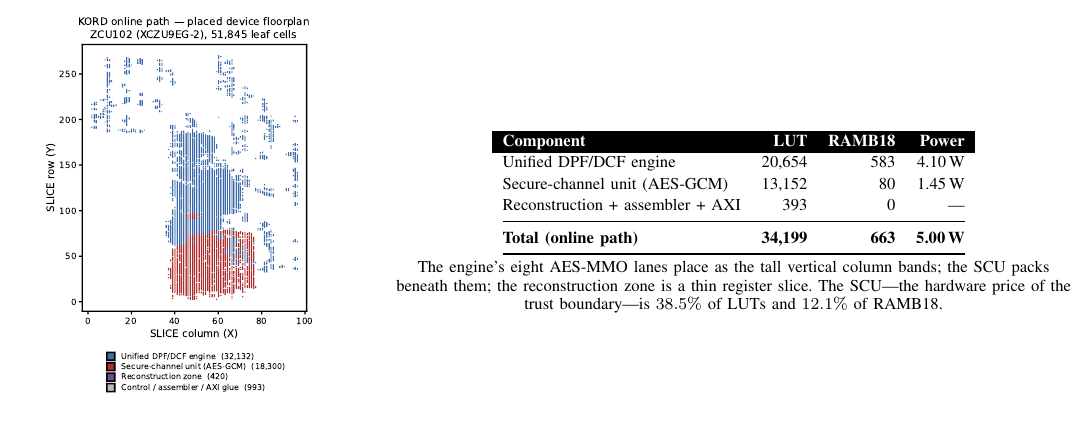}
  \caption{KORD online-path placed device floorplan and resource/power breakdown on ZCU102 (XCZU9EG-2, Vivado 2022.2). Left: the actual post-route placement---each dot is one of 51,845 leaf cells at its SLICE location, colored by component. Right: per-component LUT/RAMB18 (additive to the total) and standalone-OOC power (total power is the integrated online-path estimate, vectorless). Synthesis-estimated $f_{\max}$ is 147.0~MHz; the design is board-measured bit-exact through $\geq214.3$~MHz (Section~\ref{sec:hardware-eval}).}
  \label{fig:floorplan}
\end{figure*}

\textbf{Distributed ORAM.}
Distributed ORAM consumes a fresh DPF key for every oblivious access, so key-generation cost recurs per access. At address width $d=16$, our measured anchor, KORD's per-access generation communication is 1,179$\times$ (caliber A) to 19,907$\times$ (caliber B) below the baseline's. The supplementary material gives extrapolations to wider address spaces ($d=20$, $d=32$).

\subsection{The Hardware}
\label{sec:hardware-eval}

All hardware results follow the platform and provenance conventions of Section~VII-A. Area and power in this subsection are synthesis-estimated (Table~\ref{tab:engines}); correctness, fail-closed authentication, $f_{\max}$, and saturated throughput are board-measured on a ZCU102 and reported at the end of the subsection. We first establish key correctness independently of timing. The DPF and unified DCF engines match a software oracle bit-for-bit across 108,276 and 226,920 checks with zero errors. The full online path with the secure-channel unit passes 4,540 end-to-end checks with 60/60 authentication successes.

Cross-key pipelining lifts the shared AES array from 8.31\% single-chain utilization to 99.84\% at saturation (RTL, measured in simulation). The headline DPF-only engine (W8) runs at 204.0~MHz for 12.75~M DPF-32 keys/s on 21,540 LUT: its 4-lane issue groups let $W=8$ retire two tree levels per cycle. The unified DPF/DCF engine (W8) runs at 154.6~MHz on 20,654 LUT and issues one level to an 8-lane block per cycle, independent of mode, so it supplies 4.83~M keys/s on either an all-DPF or an all-DCF stream. A DCF level uses all eight lanes; a DPF level uses only four and idles the value lanes---the honest cost of a shared 4-domain block---so the unified engine trades the DPF-only engine's $2\times$ DPF rate for one datapath serving both functions. Mixed workloads satisfy the mode-independent capacity line $(R_{\rm DPF}+R_{\rm DCF})/4.83\mathrm{M}\leq1$ (Figure~\ref{fig:beyond}d). This per-mode rate, and the saturated utilization it assumes, are board-measured (below); the lower frequency comes from the DCF correction-word chain. The unified path has a +0.96\% LUT overhead relative to a DCF-only datapath and saves roughly half the resources of running two separate engines: 50.8\% of LUTs and 50.3\% of BRAMs (Figure~\ref{fig:beyond}c). The secure-channel unit, the hardware price of the trust boundary (implementing the hardened KORD-mac channel, an upper bound for the default keystream channel), is 38.5\% of the online system's LUT and 12.1\% of its BRAM (Figure~\ref{fig:beyond}e).

These are generation rates at the fabric's assembler boundary: a DPF-32 key share is 540~B (Section~II-A), so 12.75~M shares/s is a $\approx6.9$~GB/s egress stream per chip. Sustaining that stream off-chip requires a DMA-class interface that we have not built. KORD removes generation compute as the bottleneck; provisioning a matching delivery path remains an integration task outside this prototype.

\textbf{On-board validation.}
We programmed the online-path design (SCU, reconstruction, and unified engine, wrapped in an AXI4-Lite window at \texttt{0xA000\_0000}) onto a physical ZCU102 and drove it over the processing system's debug access port, bracketing the synthesis-estimated figures above with silicon measurements. Four results hold on the board. (i) \emph{Correctness:} 60 keys (30 DPF, 30 DCF; $\ell\in\{16,32,64\}$) reproduce the golden correction-word stream across 13,560 field checks with zero mismatches, closing the chain from RTL to silicon. (ii) \emph{Fail-closed authentication:} all four attack classes of Section~\ref{sec:security}---tampered tag, tampered ciphertext, wrong channel key, and replay of an accepted frame---are rejected with no engine output, on hardware. (iii) \emph{Frequency:} sweeping the PL clock, the full online path runs bit-exact through 214.3~MHz and fails by 250~MHz, so its real $f_{\max}$ exceeds the 147.0~MHz WNS extrapolation by at least $1.46\times$. The synthesis estimates are conservative---as expected for out-of-context timing signed off at a worst-case corner---so every rate derived from them is a lower bound. (iv) \emph{Throughput and utilization:} an on-chip request generator that keeps the engine's $M=16$ slots full reproduces the saturated regime the rate law assumes. The unified engine sustains 99.98\% lane utilization and matches its 4.83~M/s rate to within 0.1\% at every width and clock; a DPF stream on it sits at 49.99\% (one 8-lane block, half used), which is exactly the per-mode behavior that fixes the endpoint at 4.83~M/s. The DPF-only engine, measured the same way, holds 99.98\% and confirms its two-levels-per-cycle rate. The 8.31\% $\rightarrow$ 99.84\% saturation curve, previously simulation-only, is thereby confirmed on silicon.

One unified engine supplies keys for 0.88 ResNet-18 inferences per second. This rate supports $\approx10.0$ concurrent online sessions. The rate is the reciprocal of a cycle-accounted 1.139~s at 154.56~MHz. The DCF-family inventory spans depths 10--32, so this accounting is finer-grained than the endpoint rates in Table~\ref{tab:engines}. For DORAM at $d=20$ the DPF-only engine generates 20.40~M keys/s against a 40~Mbps link that drains only 833.3~K messages/s, $24\times$ more supply than the link can carry (Figure~\ref{fig:beyond}b). Provided that the key-delivery path is provisioned to match, moving generation inside the trust boundary yields modeled supply rates that exceed network demand. The link between the parties then becomes the binding constraint assumed by the online-optimization literature.

\section{Related Work}

\textbf{FSS systems and their key sources.}
Non-FSS private inference pays for every non-linear layer online~\cite{ref26,ref34,ref36,ref39}; FSS systems~\cite{ref21,ref22,ref25,ref41} shift this cost into one-time keys. Deployed systems obtain keys from a dealer or helper party~\cite{ref21,ref22,ref25,ref41,ref45,ref51} or from a point-knowing client~\cite{ref3,ref11,ref12,ref52}; neither fits an MPC pipeline, where the point is a secret-shared mask~\cite{ref16,ref48}. KORD serves this setting, exposing masks and key shares to no principal (Section~\ref{sec:security}).

\textbf{Dealerless key generation.}
Xing et al.~\cite{ref54}, our baseline, follows Floram~\cite{ref15}; we do not treat its full-tree Gen as a lower bound. Programmable DPFs~\cite{ref10}, correlation generators~\cite{ref5,ref6,ref55}, and generic 2PC~\cite{ref27} remain impractical or costlier at inference scale (Section~\ref{sec:building-blocks}); GPU batching~\cite{ref21,ref25,ref31,ref38,ref46,ref53} accelerates a route without changing its trust or interaction structure.

\textbf{Cryptographic hardware and trusted roots.}
FHE ASICs~\cite{ref28,ref29,ref30,ref42,ref43}, HE co-design~\cite{ref40}, garbled-circuit accelerators~\cite{ref35,ref44}, and OT-extension acceleration~\cite{ref33} target data-parallel kernels, not the trust model. PPMLAC~\cite{ref56} roots secret-sharing ML in attested hardware; KORD adopts only its initialization. A TEE-hosted dealer~\cite{ref13,ref32,ref47,ref50} executes loaded code with its attack surface~\cite{ref49} and remains a principal; KORD's special-purpose chips load none.

\section{Conclusion}

FSS moves the cost of private computation from the online phase into a key stream. Deployed systems pay for that stream along one of three edges of a deployment trilemma: they trust a dealer, incur interaction that grows with tree depth and security parameter, or accelerate a route without changing it. A one-time, attestable pairing of two special-purpose chips enables key generation in one round and a few bytes. Depth-serial GGM expansion leaves the hardware idle when processing one key at a time, but interleaving independent keys saturates a single AES array. This design relocates rather than eliminates the trust assumption, moving it from an online principal to minimal, auditable hardware. KORD explicitly identifies the operations that this hardware must be trusted to perform and the cost of its compromise. For FSS-based deployments whose bottleneck is the key stream itself, KORD is dealerless under a minimal hardware root.

\end{document}